\documentclass[english, authoryear]{elsarticle}
\usepackage{setspace}
\usepackage{natbib}
\usepackage{amsfonts}
\usepackage{amsmath}
\usepackage{amssymb}
\usepackage{amsthm}
\usepackage{mathrsfs}
\usepackage{graphics}
\usepackage{lscape}
\usepackage{changebar}
\usepackage{graphicx}
\usepackage{epsfig}
\usepackage{subeqnarray}

\renewcommand{\vec}[1]{\mathbf{#1}}
\newcommand{\taub}{\mbox{\boldmath$\tau$}}
\newcommand{\Deltab}{\mbox{\boldmath$\Delta$}}

\newcommand{\deriv}[2]{\frac{\partial #1}{\partial #2}}

\newcommand{\del}{\vec{\nabla}}

\begin{document}
\begin{frontmatter}
\title{Effective slip boundary conditions for arbitrary periodic surfaces:  The surface mobility tensor}

\author[Harvard]{Ken Kamrin\corref{cor1}}
\ead{kkamrin@seas.harvard.edu}
\ead[url]{http://people.seas.harvard.edu/$\sim $kkamrin/}
\author[MIT]{Martin Z. Bazant}
\author[Princeton]{Howard A. Stone}

\cortext[cor1]{Corresponding author.}

\address[Harvard]{School of Engineering and Applied Sciences, Harvard University, Cambridge, MA 01238, USA}
\address[MIT]{Department of Chemical Engineering, Massachusetts Institute of Technology, Cambridge, MA 02139, USA}
\address[Princeton]{Department of Mechanical and Aerospace Engineering, Princeton University, Princeton, NJ 08544, USA}

\begin{abstract}
In a variety of applications, most notably microfluidic design, slip-based boundary conditions have
been sought to characterize fluid flow over patterned surfaces. We
focus on laminar shear flows over surfaces with periodic height
fluctuations and/or fluctuating Navier scalar slip properties. We derive a general formula for the ``effective slip'', which describes equivalent fluid motion at the mean surface as depicted by the linear velocity profile that arises far from it. We show that the slip and the applied stress are related linearly through a tensorial mobility matrix, and the method of domain perturbation is then used to derive an approximate formula for the mobility law directly in terms of surface properties. The specific accuracy of the approximation is detailed, and the mobility relation is then utilized to address several questions, such as the determination of optimal surface shapes and the effect of random
surface fluctuations on fluid slip.
\end{abstract}
\end{frontmatter}

\section{Introduction}

With recent progress in the study and fabrication of mirofluidic devices, new interest has arisen in determining generalized forms of hydrodynamic boundary conditions \citep{stone2004, bazant2008}. Advances in lithography to pattern substrates at the micrometer and nanometer length scales have raised several questions in the modeling of fluid motions over these surfaces. For example, rather than trying to solve equations of motion for the flow at the scale of the individual corrugations of the pattern, it is appropriate to consider the bulk fluid motion (on length scales much larger than the pattern wavelength) by utilizing effective boundary conditions that characterize the flow at the surface.  For fluid being sheared horizontally over a textured surface, a basic aim is to derive a local boundary condition that can be applied along the smooth mean surface, which mimics the effects of the actual condition along the true surface. These effective conditions can be used in place of the no-slip condition to solve for macro-scale flow without the tedium of enforcing a boundary condition on a rough boundary geometry.

A standard phenomenological approach for flow over a patterned surface is to assume a Navier slip boundary condition
\begin{equation}\label{navierslipcond}
\vec{u}^s=\vec{U}-\vec{u}=b \ \deriv{\vec{u}}{n}
\end{equation}
which relates the fluid velocity $\vec{u}$ at the surface, the velocity of the surface $\vec{U}$,  and the shear strain rate normal to the mean surface, $\partial \vec{u}/\partial n$, via the slip-length $b$. This approach has been studied extensively in experiments, theoretical calculations, and simulations (for recent discussion and results see \citet{vinogradova1999,lauga2005,bocquet2007, davis09}). 

The above relationship avoids the possibility of transverse flow over a grooved no-slip surface,
perpendicular to an applied shear stress, which has been analyzed and observed in a number of studies (see
\citet{stroock2002b, ajdari2002, wang2003, stroock2002a}).  Such phenomena have motivated a tensorial version of Eq \ref{navierslipcond}, as discussed in \citet{stroock2002b} and \citet{stone2004}, which replaces the scalar slip-length $b$ with a rank-2 tensor $\vec{b}$ characterizing the surface anisotropy:
\begin{equation}
\vec{u}^s = \vec{b} \cdot (\hat{\vec{n}} \cdot \del \vec{u})  \label{eq:tensorbc}
\end{equation}
where $\hat{\vec{n}}$ is the unit normal (directed into the fluid).  \citet{bazant2008} proposed to express the tensorial slip condition in the convenient form of a mobility law, where the mean surface normal traction $\taub=\vec{T}\cdot\hat{\vec{n}}$ (for $\vec{T}$ the Cauchy stress tensor) and some mobility tensor $\vec{M}$ are used instead of the velocity gradient and tensorial slip-length, i.e.
\begin{equation}\label{tensormobil}
\vec{u}^s=\vec{M}\cdot\taub
\end{equation}
and discussed general physical constraints on $\vec{M}$ for different types of fluids and surfaces. 

The work herein analyzes and quantifies this proposed relationship for the case of Stokes flow over a broad class of weakly textured surfaces. We focus on horizontally sheared fluid over surfaces with arbitrary periodic height fluctuations, and continue the analysis later (Section \ref{navier}) to surfaces that also have non-uniform hydrophobicity.  To rigorously evaluate the properties of $\vec{u}^s$, our approach is to construct the flow from a family of \emph{analytical} solutions to the equations of motion, which are superposed as needed to satisfy the order-by-order boundary conditions on a topographically complex surface. Section \ref{basic} defines the problem and proves the validity of Eq \ref{tensormobil} by deriving the relationship directly from the Stokes equations for a no-slip surface with arbitrary periodic height fluctuations.  We then proceed in Section \ref{mobility} with the crucial and natural question of how to relate the details of the surface topography to the mobility tensor. This is considered for the case of arbitrary, small, periodic surface corrugations, where a second-order approximate formula for $\vec{M}$ is constructed from a family of Fourier series solutions to the Stokes equations and carrying out a domain-level perturbation analysis up to second-order. This result is then used to derive/compute a number of consequential results: analytical results in the case of grooved surfaces (Section \ref{stripes}), flow optimization over surfaces of fixed heterogeneity (Section \ref{opt}), and statistical results for random surfaces (Section \ref{random}).  The limitations of our approximation are also deduced and quantified through an in-depth error analysis in Section \ref{error} that carries over to the appendix.

\section{Problem setup and basics}\label{basic}
Consider a rigid, periodic surface with height
$z=H(x,y)$, with period $2L_x$ in the $x$ direction, and period
$2L_y$ in the $y$ direction.  Let $z=0$ correspond to the bottom of
the surface, so that $H(x,y)\ge 0$. Above the surface is a layer of fluid (of viscosity $\eta$) satisfying the no-slip boundary condition along the surface.  It is sheared from above, at $z\rightarrow\infty$, by a horizontal shear traction $\taub=(\tau_x,\tau_y,0)$, which induces a steady flow $\vec{u}(x,y,z)$ with pressure $p(x,y,z)$.   As a convention in this paper, we interchangeably represent horizontal vectors with 2 or 3 components depending on context --- for example $\taub$ can also represent $(\tau_x,\tau_y)$ in planar operations.


Before continuing any further, let us non-dimensionalize the problem to scale $\eta$ out of the analysis.  Let $T$ and $\mathcal{L}$ be arbitrary, fixed units of time and length respectively. Let the unit of stress be $\eta/T$.   With these units, all system variables and fields are hereby redefined to be their dimensionless counterparts, e.g.
\begin{equation}
\vec{u}\rightarrow \vec{u}\ T/\mathcal{L} \ \  ,\ \ \ \  \taub\rightarrow\taub\ T/\eta \ \ , \ \ \ \  p\rightarrow p \ T/\eta \ \ , \ \ \ \ \vec{x}\rightarrow \vec{x}/\mathcal{L} \ \ , \ \ \ \ H\rightarrow H/\mathcal{L}
\end{equation}
We assume that the Reynolds number is sufficiently small that the flow satisfies the three-dimensional Stokes equations. Under our non-dimensionalization, this gives
\begin{subeqnarray}
 \nabla^2\vec{u}=\nabla p 
\\
\nabla\cdot\vec{u}=0
\end{subeqnarray}
and the dimensionless boundary conditions are

\begin{equation}
\vec{u}(x,y,H(x,y))=\vec{0} \ \ \ , \ \ \ \  \left.\deriv{\vec{u}}{z}\right|_{z\rightarrow\infty}=\taub
\end{equation}

Far above the patterned surface, the flow must asymptote to a simple
linear flow with uniform constant pressure.  A major goal of this paper is to determine the \emph{effective slip} --- that is, the horizontal vector $\vec{u}^s$ in the asymptotic form
\begin{equation}
\vec{u}(\tilde{z}\rightarrow\infty)=\vec{u}^s+\taub \tilde{z}
\end{equation}
where
\begin{equation}
\tilde{z}=z-\langle H(x,y)\rangle
\end{equation}
measures the distance above the space-average height of the fluctuations.  By definition, a perfectly flat, no-slip surface has $\vec{u}^s=\vec{0}$ regardless of $\taub$.  When the surface has height fluctuations, $\vec{u}^s$ is usually non-zero and has a direction and magnitude depending on the stress vector $\taub$ and the surface shape.

As a consequence of the linearity of the Stokes equations, the effective slip $\vec{u}^s$ and the traction $\taub$ must be related linearly through a $2\times2$ matrix relationship of the form:
\begin{equation}\label{slip1}
\vec{u}^s=\vec{M}\cdot\taub
\end{equation}
We refer to the matrix $\vec{M}=\vec{M}(H)$ as the mobility tensor for the surface $H(x,y)$. The mobility tensor can be seen as characterizing the dependence of the net flow properties on the direction and magnitude of the applied stress.   The linearity of the relationship between $\vec{u}^s$ and $\taub$ also means that surfaces with more than two symmetry directions must have isotropic mobility.

It is straightforward to prove Eq \ref{slip1}.  For some fixed surface $H(x,y)$ consider two particular flows:  The first flow is induced by applying a unit shear traction from above pointed in the $\hat{x}$ direction, giving rise to some flow profile $\vec{U_1}(x,y,z)$ with corresponding effective slip $\vec{U_1}^s$.   The second flow is induced by a unit traction in the $\hat{y}$ direction, generating a flow profile $\vec{U_2}(x,y,z)$ with slip $\vec{U_2}^s$.  
Exploiting the linearity of the Stokes equations, the following superposition is an exact solution for the flow $\vec{u}(x,y,z)$ induced by an arbitrary shear traction $\taub=(\tau_1,\ \tau_2)$ applied from above:
\begin{equation}
\vec{u}(x,y,z)=\tau_1\vec{U_1}(x,y,z)+\tau_2\vec{U_2}(x,y,z)
\end{equation}
Consequently, the effective slip $\vec{u}^s$ arising from the shear traction $\taub$ is
\begin{equation}
\vec{u}^s=\tau_1\vec{U_1}^s+\tau_2\vec{U_2}^s=\big[\vec{U_1}^s  |   \vec{U_2}^s \big]\cdot\taub
\end{equation}
The matrix $[\vec{U_1}^s | \vec{U_2}^s]$ is our mobility matrix $\vec{M}$.

\begin{figure}
\begin{center}
(a)\epsfig{file=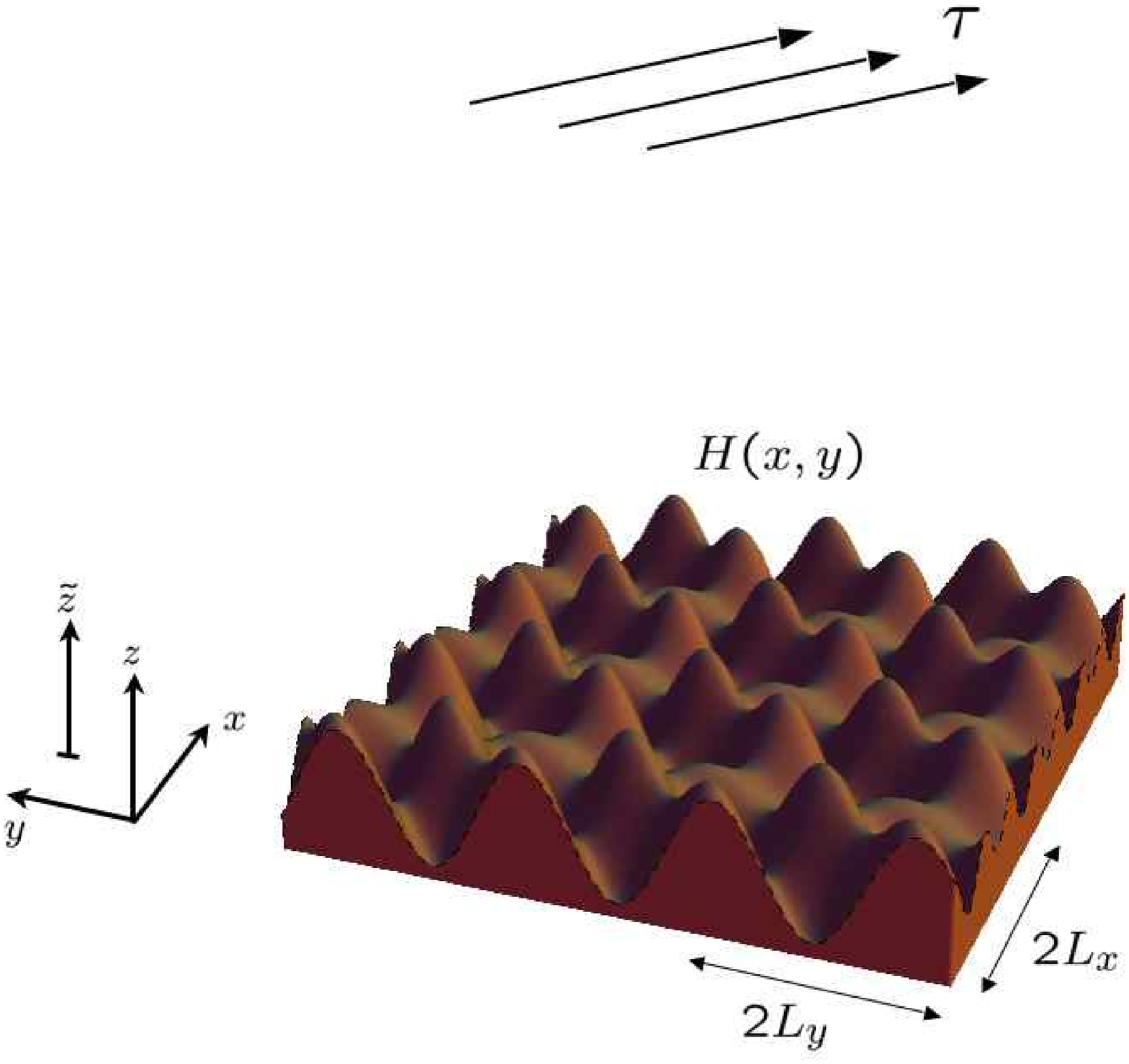, width=2.8in,clip} \ \ \ \ \  \ (b) \epsfig{file=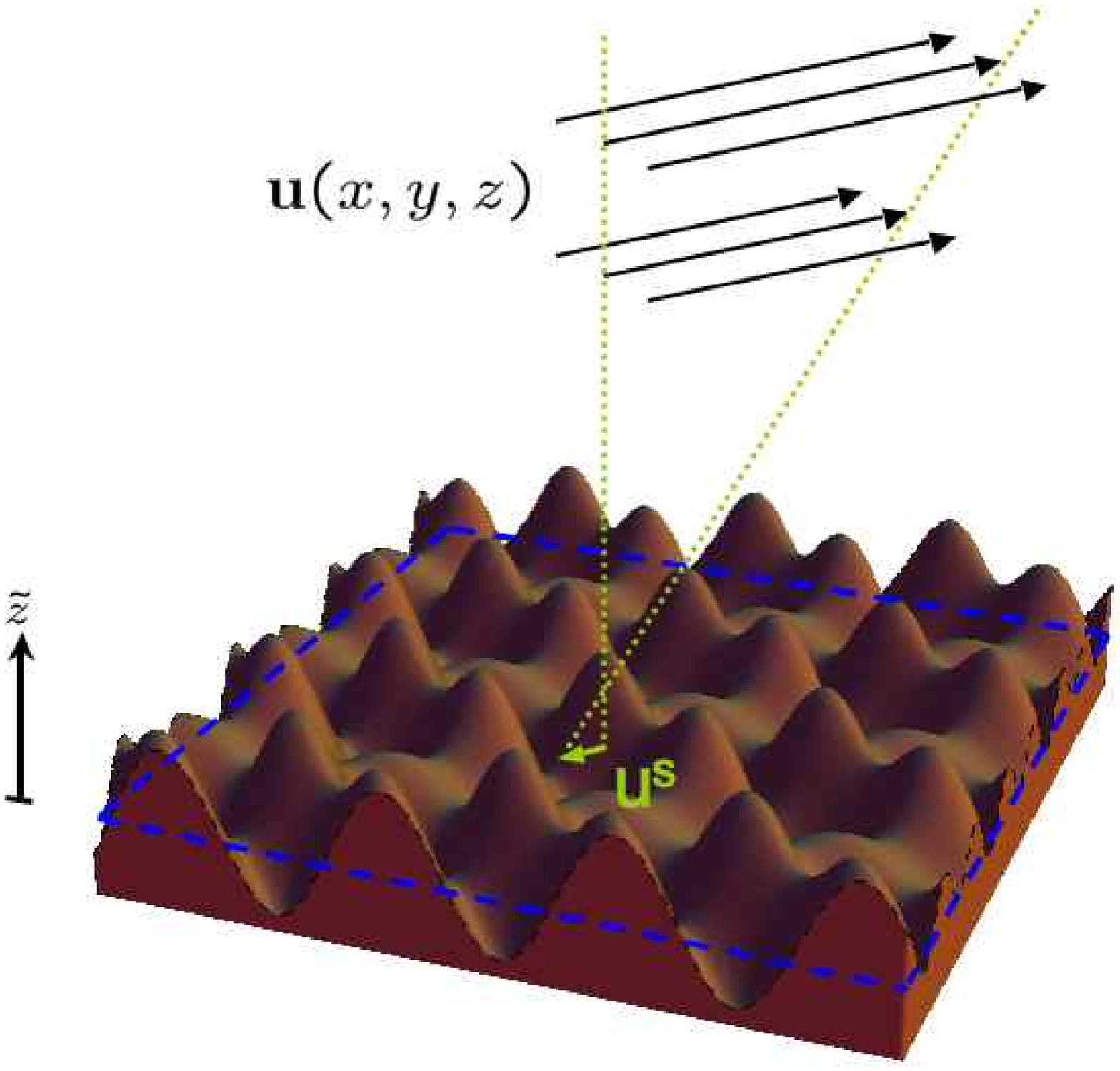, width=2.47in,clip}
\caption{(a) Setup for the problem.  A fluid is horizontally sheared with stress $\taub$ at a height of $\tilde{z}=\infty$ above a surface with arbitrary periodic height fluctuations $H(x,y)$. (b) The induced flow profile becomes horizontal and linear in $\tilde{z}$ at large heights, but is more complicated closer to the surface.  Our interest is to determine the effective slip $\vec{u}^s$, pictured above, which corresponds to the extra velocity one would obtain when extrapolating the linear portion down to the mean surface height $\tilde{z}=0$. }\label{setup:fig}
\end{center}
\end{figure}

\section{Computing the mobility tensor}\label{mobility}
An exact formula for $\vec{M}$ in terms of $H(x,y)$ seems very complicated to obtain, and unlikely to have a tractable form.  In the following sections,  we derive an approximate formula for the $\vec{M}$ tensor in the case of small height fluctuations, i.e. 
\begin{equation}
H(x,y)=\epsilon h(x,y)
\end{equation}
for $\epsilon$ some small, dimensionless number.  

It is difficult to ascribe a single physical description to $\epsilon$ at the outset, since there are an infinite number of possible size scales that could be extracted from an \emph{arbitrary} periodic surface.  Which dimensionless size must be small for the approximation to be valid? To approach this question systematically, our strategy is to obtain the approximate solution for $\vec{M}$ using domain perturbation theory treating $\epsilon$ as a small parameter, and then perform an in-depth error analysis to determine how the error of the approximation depends on $h(x,y)$.  Once the error is quantified in this fashion, it can be used to single out a particular definition of $\epsilon$ that is ``ideal'' in the sense that the relative approximation error depends solely on $\epsilon$ and no other surface properties.

We now summarize the major result of this paper.  Let $\hat{h}(m,n)$ compose the set of Fourier coefficients corresponding to $h(x,y)$
\begin{equation}
h(x,y)=\sum_{m,n}\hat{h}(m,n)e^{i(k_m x+k_n y)}
\end{equation}
 where $k_m=m\pi/L_x$ and $k_n=n\pi/L_y$ are corresponding wavenumbers.  Then the effective slip velocity obeys

\begin{equation}\label{sliprelation}
\vec{u}^s=-\epsilon^2\tilde{\vec{M}}(h)\cdot\taub +O(\epsilon^3)
\end{equation}
where the function $\tilde{\vec{M}}(h)$ is defined by
\begin{equation}\label{slipmatrix}
\tilde{\vec{M}}\big(h\big)=
\begin{pmatrix}\displaystyle
\sum_{(m,n)\neq \vec{0}}\frac{2k_m^2+k_n^2}{\sqrt{k_m^2+k_n^2}}\left|\hat{h}(m,n)
\right|^2 & &
\displaystyle\sum_{(m,n)\neq \vec{0}} \frac{k_mk_n}{\sqrt{k_m^2+k_n^2}}\left|\hat{h}
(m,n)\right|^2
\\
\displaystyle\sum_{(m,n)\neq \vec{0}} \frac{k_mk_n}{\sqrt{k_m^2+k_n^2}}\left|\hat{h}
(m,n)\right|^2 & &
\displaystyle\sum_{(m,n)\neq \vec{0}}\frac{k_m^2+2k_n^2}{\sqrt{k_m^2+k_n^2}}\left|
\hat{h}(m,n)\right|^2
\end{pmatrix}
\end{equation}
Hence, $-\epsilon^2\tilde{\vec{M}}$ is a second-order approximation to the true mobility matrix $\vec{M}$.
  
First, observe that the mobility is an $O(\epsilon^2)$ effect for height fluctuations of size $O(\epsilon)$.  A second-order leading perturbation term was also observed in \cite{stroock2002b} for the effective slip of pressure-driven Poiseuille flow along sinusoidally grooved walls.  Furthermore, the leading-order mobility matrix is  {\it symmetric} regardless of  $h(x,y)$ (within the broad range of validity of the analysis, quantified below).  This result was suggested, though not proved, in \cite{bazant2008}, where the notion of symmetric mobility was supported by a statistical diffusion argument and treated as an example of the commonly used Onsager-Casimir relations for near-equilibrium linear response.  The symmetry of $\tilde{\vec{M}}$ requires that its eigenvectors are orthogonal, and likewise a general surface $H$ should have two orthogonal directions along which the shear stress and slip direction align.  If stress is applied in a different direction, the apparent slip can have lateral components that transport fluid transverse to the direction of the stress traction, as in \cite{stroock2002b} and \cite{stone2004}.

It is also important to note that with the definitions used here, Eqs \ref{sliprelation} and \ref{slipmatrix} give a negative-definite mobility.  That is, the effective slip should point away from the traction vector (see Figure \ref{setup:fig}(b)), which reflects the notion that surface fluctuations cause flow \emph{resistance} compared to a flat surface of the same mean height. This outcome can also be seen as directly related to our choice of placing the origin for $\tilde{z}$ at the mean height of the pattern.  In studies such as \cite{wang2003}, the origin is placed at the peak of the height fluctuations, which redefines the slip vector in a way that ensures $\taub\cdot\vec{u}^s\ge 0$, and equivalently redefines the mobility tensor to be positive-definite.  Positive mobility is desirable from an intuitive standpoint because it more closely represents an object slipping on a dissipative, ``passive surface'' \citep{bazant2008}.  However, for our purposes, we find it computationally easier to perform our analyses with the mean surface height selected as the origin.  Other effective slip studies such as \cite{stroock2002b} have also found benefit in setting the origin at the surface mean.  The remainder of this section details the derivation of Eqs \ref{sliprelation} and \ref{slipmatrix}.

\subsection{Perturbation expansion}\label{pert_series}
Our approach utilizes the method of domain perturbations (see \citet{hinch1991}).  \citet{miksis1994} have also used this technique to study surface flows over small-fluctuation surfaces.  Their focus was on two-dimensional flows over a surface $H(x)$, and the bulk flow behavior was approximated using asymptotic matching.  Here, we study the three dimensional problem with bulk behavior solved analytically, and carry out the domain perturbation analysis to higher order as necessary to reveal the tensorial properties of the effective slip at the surface.

To begin, we let the velocity and pressure be represented by the perturbation series
\begin{subeqnarray}
\vec{u}(x,y,z)=\vec{u_0}(x,y,z)+\epsilon \vec{u_1}(x,y,z)
+\epsilon^2 \vec{u_2}(x,y,z)+O(\epsilon^3)
\\
p(x,y,z)=p_0(x,y,z)+\epsilon p_1(x,y,z) +\epsilon^2
p_2(x,y,z)+O(\epsilon^3).
\end{subeqnarray}

Inserting these into the Stokes equations
and grouping same-order terms, it is immediately clear that the Stokes
equations must be upheld at each individual order. Using the method of domain perturbations, the expression of
the boundary condition along the periodic surface becomes much easier
to deal with by expanding in a Taylor series about $z=0$:
\begin{align*}
\vec{u}(x,y,\epsilon h(x,y))=\vec{0}&=\vec{u_0}(x,y,0)+\epsilon h(x,y)\left.
\deriv{\vec{u_0}}{z}\right|_{z=0}+\frac{(\epsilon h(x,y))^2}{2}\left.\deriv{^2\vec{u_0}}
{z^2}\right|_{z=0}
\\
&+ \epsilon\left(\vec{u_1}(x,y,0)+\epsilon h(x,y)\left.\deriv{\vec{u_1}}{z}\right|
_{z=0}\right)+\epsilon^2\vec{u_2}(x,y,0)+O(\epsilon^3).
\end{align*}
This expression is equivalent to three separate boundary conditions for the first 
three orders:
\begin{subeqnarray}
\vec{u_0}(x,y,0)&=&\vec{0}
\\
\vec{u_1}(x,y,0)&=&-h(x,y)\left.\deriv{\vec{u_0}}{z}\right|_{z=0} \slabel{bc1}
\\
\vec{u_2}(x,y,0)&=&-h(x,y)\left.\deriv{\vec{u_1}}{z}\right|_{z=0}  -\frac{h(x,y)^2}{2}\left.
\deriv{^2\vec{u_0}}{z^2}\right|_{z=0}. \label{sbc2}
\end{subeqnarray}
The traction condition at $z\rightarrow \infty$ is first-order in magnitude, which implies the 
following three boundary conditions:
\begin{equation}
\left.\deriv{\vec{u_0}}{z}\right|_{z=\infty}=\taub \ \   \ \ \ \ \ \left.\deriv{\vec{u_1}}{z}\right|_{z=\infty}=\vec{0} \ \   \ \ \ \ \ \left.\deriv{\vec{u_2}}{z}\right|_{z=\infty}=\vec{0}.
\end{equation}

\subsection{Solving for horizontally periodic Stokes flow}

By inspection, we find that the zeroth-order solution is standard simple shear:
\begin{align}
\vec{u_0}(x,y,z)=\taub z  \ \ \ \ \ \ \ \ p_0(x,y,z)=K_0.
\end{align}
where $K_0$ is a constant.
To compute each of the higher-order terms, we need an analytic general
solution to the Stokes equations that satisfies the $z\rightarrow\infty$
boundary condition. From such a general solution, we can fit the more
complicated $z=0$ boundary conditions. Due to the horizontal
periodicity of the surface, it follows that any flow solution be
similarly periodic in the $x$ and $y$ directions:
\begin{align}\label{fourier}
&\vec{u}(x,y,z)=\sum_{m,n}\vec{a}(m,n,z)e^{i(k_m x+k_n y)}
\\
&p(x,y,z)=\sum_{m,n}b(m,n,z)e^{i(k_m x+k_n y)}.
\end{align}
Substituting these forms into the Stokes equations, a somewhat lengthy
general solution for $\vec{a}$ and $b$ can be found (see Appendix \ref{stokes}), which ultimately
depends on three sets of undetermined coefficients.  Whenever $m$ and $n$
are non-zero, $\vec{a}$ and $b$ both decay exponentially in $z$ as
expected, which guarantees satisfaction of the upper boundary condition.

\subsection{First-order term}\label{first}
To solve for the first-order flow, the undetermined coefficients are solved
term-by-term by enforcing the bottom boundary condition (Eq \ref{bc1}), which now reads 
\begin{equation}
\vec{u}_1(x,y,0)=-h(x,y)\taub.
\end{equation}
The result is
\begin{align}
u_1(x,y,z)&=-\hat{h}(0,0)\tau_x+\sum_{(m,n)\neq \vec{0}}\hat{h}(m,n)e^{-
\sqrt{k_m^2+k_n^2}z}\left(\frac{k_m^2\tau_x+k_mk_n\tau_y}{\sqrt{k_m^2+k_n^2}}z-
\tau_x\right)e^{i(k_m x+k_n y)}\label{firstordersolnu}
\\
v_1(x,y,z)&=-\hat{h}(0,0)\tau_y +\sum_{(m,n)\neq \vec{0}}\hat{h}(m,n)e^{-
\sqrt{k_m^2+k_n^2}z}\left(\frac{k_n^2\tau_y+k_mk_n\tau_x}{\sqrt{k_m^2+k_n^2}}z-
\tau_y\right)e^{i(k_m x + k_n y)}\label{firstordersolnv}
\\
w_1(x,y,z)&=\sum_{(m,n)\neq \vec{0}}\hat{h}(m,n)ie^{-\sqrt{k_m^2+k_n^2}z}(k_m 
\tau_x+k_n\tau_y)z e^{i(k_m x + k_n y)}\label{firstordersolnw}
\\
p_1(x,y,z)&=K_1+\sum_{(m,n)\neq \vec{0}}\hat{h}(m,n)2i e^{-\sqrt{k_m^2+k_n^2}z}
(k_m \tau_x+k_n\tau_y)e^{i(k_m x + k_n y)}, \label{firstordersolnp}
\end{align}
where $\vec{u_1}=(u_1,v_1,w_1)$ and $K_1$ is an arbitrary constant.  The constant terms in the $u_1$ and $v_1$ 
expansions grow proportionally to the average surface height $\hat{h}(0,0)$, and provide a first-order isotropic slip contribution, as all other terms decay exponentially as $z\rightarrow\infty$.

\subsection{Second-order term}\label{second}
To determine the second-order solution, we take the general solution from Appendix \ref{stokes} and enforce the $z=0$ boundary condition, which now reads
\begin{equation}
\vec{u_2}(x,y,0)=-h(x,y)\left.\deriv{\vec{u_1}}{z}\right|_{z=0}
\end{equation}
where
\begin{align}
\left.\deriv{u_1}{z}\right|_{z=0}&=\sum_{(m,n) \neq \vec{0}}\frac{2k_m^2\tau_x
+k_n^2\tau_x+k_mk_n\tau_y}{\sqrt{k_m^2+k_n^2}}\hat{h}(m,n)e^{i(k_m x + k_n y)}
\\
\left.\deriv{v_1}{z}\right|_{z=0}&=\sum_{(m,n)\neq \vec{0}}\frac{2k_n^2\tau_y
+k_m^2\tau_y+k_mk_n\tau_x}{\sqrt{k_m^2+k_n^2}}\hat{h}(m,n)e^{i(k_m x + k_n y)}
\\
\left.\deriv{w_1}{z}\right|_{z=0}&=\sum_{(m,n)\neq \vec{0}}i(k_m\tau_x+k_n\tau_y)
\hat{h}(m,n)e^{i(k_m x + k_n y)}
\end{align}
as follows from Eqs \ref{firstordersolnu}, \ref{firstordersolnv}, and \ref{firstordersolnw}.

Since we are stopping the perturbation analysis beyond
$O(\epsilon^2)$, there is no need to compute the full
solution for $\vec{u_2}$--- one need only determine what the constant
term is in its  Fourier series solution, Eq \ref{fourier}.  This term will be
the only one that does not vanish as $z\rightarrow\infty$, and
represents the effective slip. Note that the constant term in the
Fourier expansion of $\vec{u_2}(x,y,z)$ is necessarily equal to the
horizontal average $\langle \vec{u_2}\rangle$ at any fixed $z$, since
all other terms average to $0$ over the period.  Hence,
we can compute this constant slip term by averaging over the lower boundary condition,
\begin{equation}
\vec{u_2}^s=\langle \vec{u_2}(x,y,0) \rangle=-\left\langle h(x,y)\left.
\deriv{\vec{u_1}}{z}\right|_{z=0}\right\rangle.
\end{equation}
Since the term on the right-hand side is the average of the product of two periodic 
functions of equal periodicity, we obtain $\vec{u_2}^s$ from the product of the two 
Fourier expansions. For any coefficients $\alpha(m,n)$, recall the general rule
\begin{align}
\left\langle h(x,y)\sum_{m,n}\alpha(m,n)\hat{h}(m,n)e^{i(k_m x+k_n y)}\right
\rangle&=\sum_{m,n}\alpha(m,n)\hat{h}(m,n)\hat{h}(-m,-n) \nonumber
\\
&=\sum_{m,n}\alpha(m,n)\left|\hat{h}(m,n)\right|^2.
\end{align}
Applying this rule to the given boundary conditions, we uncover the tensor $\tilde{\vec{M}}(h)$ from Eq \ref{slipmatrix} giving
\begin{equation}
\vec{u_2}^s=-\tilde{\vec{M}}(h)\cdot\taub
\end{equation}
Combining the constant terms from the first- and second-order flows with the zeroth-order shear flow, we obtain the limiting behavior 
\begin{equation}
\vec{u}(z\rightarrow\infty)=\taub z-\epsilon\underbrace{\hat{h}(0,0)}_{=\langle h(x,y)\rangle}\taub-\epsilon^2\tilde{\vec{M}}(h)\cdot\taub + O(\epsilon^3)
\end{equation}
By switching to the variable $\tilde{z}$, the first-order contribution cancels yielding 
\begin{equation}
\vec{u}(\tilde{z}\rightarrow\infty)=\taub\tilde{z}-\epsilon^2\tilde{\vec{M}}(h)\cdot\taub + O(\epsilon^3)
\end{equation}
and the slip formula (Eq \ref{sliprelation}) immediately follows.

\section{Error analysis}\label{error}
\subsection{The need to compute error bounds}
The significance of the slip approximation (Eq \ref{sliprelation}) is tied directly to the error of truncating the perturbation expansion at second order.  It is of critical importance that we quantify this error to correctly interpret the mobility matrix approximation $\vec{M}(\epsilon h)\approx-\epsilon^2\tilde{\vec{M}}(h)$.  

An example is instructive here. Consider a surface of shallow, square grooves--- that is, $H(x,y)=\epsilon h(x,y)$ where

\begin{equation}
h(x,y)=\left\{
\begin{array}{ll}
1 & \text{if $2k<x\le 2k+1$ for integer $k$}
\\
0 & \text{if $2k+1<x\le 2k+2$ for integer $k$}.
\end{array}
\right.
\end{equation}
To use the mobility matrix approximation for this surface (Eq \ref{slipmatrix}), one computes the Fourier coefficients of the above and finds that $|\hat{h}(m,n)|$ decays like $|1/m|$.  Likewise, the summand in each of the diagonal entries of $\tilde{\vec{M}}$ must decay as $|1/m|$, and consequently $\tilde{M}_{11}$ and $\tilde{M}_{22}$ are both infinite.  This prediction for the mobility is clearly erroneous, as it suggests the effective slip velocity should be infinite (regardless of the choice of $\epsilon>0$), which contradicts known results in \cite{wang2003} and basic intuition.

The reason for this failure is because this particular surface introduces an infinite amount of approximation error.  In view of Eq \ref{sliprelation}, the error in $\vec{u}^s$ must grow as $O(\epsilon^3)$, which means the error can be expressed as $\epsilon^3 C(h,\taub)$ for some unknown function $C$.  However, as the example has just demonstrated, certain surfaces $h$ cause $C(h,\taub)$ to become large or infinite, which overwhelms the fact that the prefactor is $\epsilon^3$ small.  Hence, in the upcoming subsection, we bound $|C(h,\taub)|$ by some known form $|\taub|\  \text{Err}(h)$, so that
\begin{equation}\label{fullexpress}
\left|\vec{u}^s-\left(-\epsilon^2\tilde{\vec{M}}(h)\cdot\taub\right)\right|\le\epsilon^3|\taub|\ \text{Err}(h)
\end{equation}
Once the function $\text{Err}$ is determined, we can affirm that the mobility matrix approximation $\vec{M}\approx -\epsilon^2\tilde{\vec{M}}$ is valid for any surface where the error bound is small compared to the approximate slip, i.e.
\begin{equation}\label{small}
\text{Relative Error Bound}=\frac{\epsilon^3 \text{Err}(h)|\taub|}{|-\epsilon^2\tilde{\vec{M}}(h)\cdot\taub|}\approx \frac{\epsilon^3\text{Err}(h)}{\epsilon^2|\tilde{\vec{M}}(h)|}\ll 1
\end{equation}
Such a relation gives an a priori method for ruling out surface shapes for which the slip matrix approximation should fail.

Below, we show that (\ref{small}) can also be used to determine an ideal definition of $\epsilon$ in terms of properties of the surface.  Up to this point, we have described $\epsilon$ merely as a ``small dimensionless number'', and $h(x,y)$ is defined as $H(x,y)/\epsilon$.  A specific formula for $\epsilon$ as a function of the surface $H(x,y)$ is physically desirable, wherein the size of $\epsilon$ immediately relates to the accuracy of the approximation.

\subsection{Bounding the approximation error}\label{bounderror}

The second-order flow solution is a superposition of analytical solutions to the Stokes equations.  Hence, the PDEs are always satisfied exactly. Likewise, the error of the slip approximation arises solely from error in fitting the boundary condition $\vec{u}(x,y,\epsilon h(x,y))=\vec{0}$.  

Appendix \ref{derivation} constructs a bound on this error, which leads to the following bound on the error of the second-order slip approximation:
\begin{equation}\label{errorwitheps}
\left|\vec{u}^s-\left(-\epsilon^2\tilde{\vec{M}}(h)\cdot\taub\right)\right|\le \epsilon^3 \kappa \ h_M |\taub| \big(h_M\ \left|\nabla\nabla h\right|_M+|\nabla h|_M^2\big)
 \end{equation}
where we use the subscript $_M$  to denote a function's maximum value over all $x$ and $y$.  The constant $\kappa$ is dimensionless and independent of $\taub$ and the surface shape; a first calculation gives $\kappa<55$. Based on the analysis in Appendices \ref{derivation} and \ref{r1r2}, we speculate it is not possible to bound the error in a general fashion using fewer than the first two derivatives of $h$.  While this decrees a qualitative level of tightness, it is clear from the appendices that quantitatively tighter bounds could be written using non-local norms and surface integrals.  We prefer the above, as it is expressible using basic quantities.  Recent work on bounding the error of Reynold's lubrication approximation for thin-channel Stokes flows \citep{wilkening2009} has found some similar dependences on the norms of the derivatives of the surface.

Equation \ref{errorwitheps} reveals that the error is unbounded for any surface where $\left|\nabla\nabla h\right|_M$ or $|\nabla h|_M$ $\rightarrow\infty$.  This result gives the following crucial limitation:  \textbf{The approximation $\vec{M}(\epsilon h)\approx -\epsilon^2\tilde{\vec{M}}(h)$ should not be used for surfaces with corners or vertical slopes.}
This point, which might not be surprising to those familiar with domain perturbation techniques, provides an explanation for why the square grooves example of the last subsection failed.  The mobility matrix formula can only be reliably applied to $C^2$ functions $h(x,y)$.  Note that the error does not depend at all on the higher-order derivatives of $h$ (order $\ge 3$), and as such the approximation can be accurate even if the higher derivatives of the surface shape are discontinuous.

Equation \ref{errorwitheps} can also be used to read off a formula for the function $\text{Err}$ that was sought after in Eq \ref{fullexpress}:
\begin{equation}\label{final_err}
\text{Err}(h)=\kappa \ h_M \big(h_M\ \left|\nabla\nabla h\right|_M+|\nabla h|_M^2\big)
\end{equation}
Observing Eqs \ref{final_err} and \ref{slipmatrix}, we note that the relative error bound (Eq \ref{small}) can be expressed solely in terms of the actual surface $H(x,y)$, and completely independent of $\epsilon$, since all powers of $\epsilon$ can be brought into the arguments of the functions $\tilde{\vec{M}}$ and $\text{Err}$, and eliminated by using $H=\epsilon h$:
\begin{align}\label{relerrorbound}
\text{Relative Error Bound}=\frac{\epsilon^3 \text{Err}(h)}{\epsilon^2|\tilde{\vec{M}}(h)|}=\frac{\text{Err}(H)}{|\tilde{\vec{M}}(H)|}
\end{align}

This last expression is a dimensionless ratio determined entirely by the given surface $H(x,y)$. While it may not be simple to interpret physically for a general surface, its direct connection to the error suggests an ideal formula for choosing $\epsilon$ given $H(x,y)$ would be

\begin{equation}\label{epsdef}
\epsilon_{ideal}\equiv\frac{\text{Err}(H)}{ |\tilde{\vec{M}}(H)|}=\frac{\kappa \ H_M \big(H_M\ \left|\nabla\nabla H\right|_M+|\nabla H|_M^2\big)}{\left| 
\begin{pmatrix}\displaystyle
\sum_{(m,n)\neq \vec{0}}\frac{2k_m^2+k_n^2}{\sqrt{k_m^2+k_n^2}}\left|\hat{H}(m,n)
\right|^2 & &
\displaystyle\sum_{(m,n)\neq \vec{0}} \frac{k_mk_n}{\sqrt{k_m^2+k_n^2}}\left|\hat{H}
(m,n)\right|^2
\\
\displaystyle\sum_{(m,n)\neq \vec{0}} \frac{k_mk_n}{\sqrt{k_m^2+k_n^2}}\left|\hat{H}
(m,n)\right|^2 & &
\displaystyle\sum_{(m,n)\neq \vec{0}}\frac{k_m^2+2k_n^2}{\sqrt{k_m^2+k_n^2}}\left|
\hat{H}(m,n)\right|^2
\end{pmatrix}
\right|}
\end{equation}
The relative approximation error can never exceed $\epsilon_{ideal}$, and hence all surfaces with a small value for $\epsilon_{ideal}$ are  well-described by the mobility formula Eq \ref{slipmatrix}.  If we let $\epsilon=\epsilon_{ideal}$ in Eq \ref{fullexpress}, and consequently  $h(x,y)=H(x,y)/\epsilon_{ideal}$, then the slip relation can be rewritten as
\begin{equation}\label{errorideal}
\left|\vec{u}^s-\left(-\epsilon_{ideal}^2\tilde{\vec{M}}(h)\cdot\taub\right)\right|\le\epsilon_{ideal}^3 \ |\tilde{\vec{M}}(h)|\ |\taub|
\end{equation}

The definition of $\epsilon_{ideal}$ reduces to a simple, expected form in canonical cases.  For example, consider the surface
\begin{equation}
H(x,y)=a \sin(x/b)
\end{equation}
for constants $a$ and $b$.  We expect that the relevant dimensionless quantity in this case should be $a/b$, in the sense that the slip approximation over this surface should improve as the height to wavelength ratio decreases to zero.  Evaluating (\ref{epsdef}), the relevant small quantity according to the error bound is $\epsilon_{ideal}=(8\kappa/\sqrt{5})(a/b)$, which is proportional to $a/b$ as expected.

To conclude this discussion on error analysis, we emphasize the major points. Most importantly, the approximate slip given by Eqs \ref{sliprelation} and \ref{slipmatrix} can be inaccurate for certain surfaces.  Approximation error grows with the maximal curvature and maximal slope of the surface.  To determine the accuracy of the approximation, there are two mathematically identical representations that have been discussed (Eqs \ref{errorwitheps} and \ref{errorideal}).  The representation shown in Eq \ref{errorwitheps} is most applicable when the user wishes to decide the value of $\epsilon$ based on a loose observation of the characteristic surface height.  Such an approach is useful when computing the mobility properties of a broad family of surfaces, all with similar characteristic height $\epsilon$.  To survey the range of surfaces, one need only vary the scaled surface $h$, taking care to ensure the range includes only those $h$ shapes that keep Eq \ref{errorwitheps} within one's error tolerance.  On the other hand we also present the error bound in a second form, that of Eq \ref{errorideal}, which is in terms of a specific definition of $\epsilon$ extracted directly from the surface $H(x,y)$. This error representation is best when there is only one surface $H$ to be studied, since $\epsilon_{ideal}$ is in the truest sense the small parameter of interest, which immediately determines the relative accuracy of the approximation.  However, if a comparison over many surfaces is desired, the need to recompute $\epsilon_{ideal}$ for each surface $H$ can complicate the analysis.

\section{Mobility over surfaces varying in only one direction}\label{stripes}

Our main result above (Eq \ref{slipmatrix}) is useful to reveal general properties of the effective slip and, within the error bounds, to allow its computation for an arbitrary surface.  However, the various series for the elements of $\tilde{\vec{M}}$ can be cumbersome to evaluate.  Of course, it is easy to evaluate them for surfaces with simple sinusoidal perturbations, which have Fourier series truncated after a small number of terms, but  this offers little analytical insight into the dependence of the mobility tensor on the shape of the surface. To more clearly see this dependence, this section focuses on a simpler family of surfaces.

Observing Eq \ref{slipmatrix}, we note that the formula reduces significantly for the class of surfaces varying in only one direction.  Letting the variation direction be $x$, these surfaces appear as parallel groove patterns with shape $h(x)$ and mobility

\begin{equation}\label{grooves}
\vec{M}\approx -\epsilon^2 \tilde{\vec{M}}\big(h\big)=
-\epsilon^2 \beta
\begin{pmatrix}\displaystyle
 2  & &
0
\\
0 & &
1
\end{pmatrix} \  , \ \ \ \  \text{for} \ \ \ \ \ \beta=2\sum_{m=1}^{\infty} k_m  \left|\hat{h}(m)\right|^2
\end{equation}
This shows that the slip length, which measures distance from the mean surface height to the equivalent no-slip plane, is generally twice larger for perpendicular versus parallel shearing.  For anisotropic Stokes flow, factors of two like this are not surprising--- it is reminiscent of similar results for striped pipes \citep{lauga2003} and the classical result that a rod sediments twice as fast in creeping flow if aligned vertically, rather than horizontally \citep{batchelor1970}. 

\begin{figure}
\begin{center}
\epsfig{file=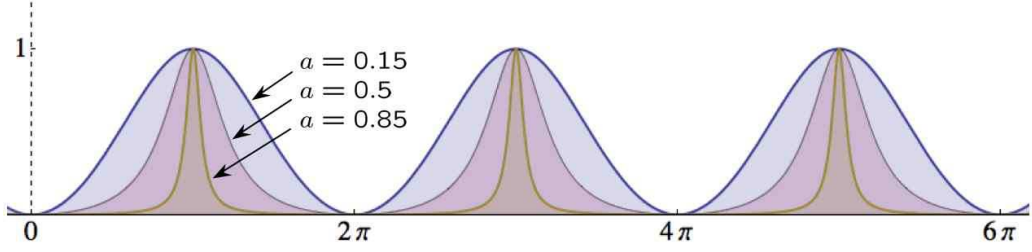, width=5.4in} 
\caption{The surface shape $h(x)$, defined in Eq \ref{groove_family}, under different values of the parameter $a$.}\label{groove_a_pic}
\end{center}
\end{figure}

Let us evaluate $\vec{M}$ in closed form for a nontrivial subset of grooved surfaces, whose Fourier series have an infinite number of terms, and vary continuously in shape from sinusoidal to sharply peaked.  Consider surface shapes $h(x)$ of the type

\begin{equation}\label{groove_family}
h(x)=\frac{\phi(x)-\phi_{min}}{\phi_{max}-\phi_{min}} \ \ \ , \ \ \ \ \phi(x)=\frac{1}{1+a^2+2a\cos x}
\end{equation}
and the unit of length is $\mathcal{L}=L_x/\pi$ for simplicity, so that $k_m=m$.  Regardless of the parameter $0<a<1$,   $h(x)$ has a maximum of $1$, and a minimum of $0$.  As shown in Figure \ref{groove_a_pic}, $a$ controls the shape of the surface. For $a \ll 1$, the surface height is a small perturbation from a single-mode sinusoid. As $a \to 1$, the surface develops wide deep valleys around the minimum height at $x=2n\pi$ separated by tall, narrow peaks at $x=(2n+1)\pi$. 

The Fourier coefficients of $h(x)$ can be obtained directly from those of $\phi(x)$. Set $z=e^{ix}$ to obtain a rational function $f(z)$, separate into partial fractions,
and expand in geometric series (since $|az|=|a/z|<1$) to obtain the Laurent series of $f(z)$, which equals the desired Fourier series on the unit circle:
\begin{eqnarray*}
\phi(x) &=& \frac{z}{(az+1)(z+a)}  = \frac{1}{1-a^2} \left(\frac{1}{1+az} - \frac{a}{z} \cdot \frac{1}{1+a/z} \right)  \\
&=& \frac{1}{1-a^2} \left( 1 + \sum_{m=1}^\infty (-a)^m (z^m+z^{-m}) \right) = \sum_{m=-\infty}^{\infty} \frac{(-a)^{|m|} e^{imx} }{1-a^2}
\end{eqnarray*}
Consequently,
\begin{equation}
\hat{h}(m\neq0) = \frac{1}{\phi_{max} - \phi_{min}}\frac{(-a)^{|m|}}{1-a^2}=\frac{a^2-1}{4a}(-a)^{|m|}
\end{equation}
By Eq \ref{grooves},
\begin{equation}
\beta=  2 \sum_{m=1}^\infty \left(\frac{a^2-1}{4a}\right)^2 m (-a)^{2m} = \left(\frac{a^2-1}{4a}\right)^2 a\ \frac{d}{da} \sum_{m=0}^\infty  a^{2m}=\frac{1}{8}\end{equation}

We observe that for this specific family of surfaces, $\beta$ does not vary with $a$.  However, the observed slip properties will vary, since the average surface height, at which the slip calculation applies, does change with $a$. To understand the slip behavior more clearly, we use Eq \ref{grooves} to compute a second-order approximation to the location of the equivalent no-slip plane (denoted $z^{N.S.}$, i.e. the height of the plane having zero mean velocity) corresponding to each $a$ under both parallel and transverse shear.  The mean surface height obeys $\epsilon\langle h \rangle=\epsilon(1-a)/2$, and consequently, for any $\epsilon$ and unit applied shear stress, we have to second-order
\begin{subeqnarray}
z^{N.S.}_{ \  \parallel}\approx\epsilon\frac{1-a}{2}+\frac{\epsilon^2}{8} \slabel{zpar}
\\
z^{N.S.}_{\ \perp} \approx\epsilon\frac{1-a}{2}+ \frac{\epsilon^2}{4} \slabel{zperp}
\end{subeqnarray}
This qualitatively states that as $a$ increases, the effective no-slip plane descends (regardless of the shear direction).  In physical terms, the fluid content in the groove pattern increases with $a$, which increases the lubrication of the surface.  The directionality of the surface comes in at second-order, where the added flow resistance of shearing transverse to the grooves is apparent.

As $a$ increases, it is important to keep in mind the accuracy of the approximation.  The second derivative of $h$ and the square of the first derivative both diverge as $(1-a)^{-2}$ as $a\to 1$.  Hence, Eq \ref{errorwitheps} implies the error bound on $z^{N.S.}$ must also diverge as $a\to 1$,
\begin{equation}\label{groove_bound}
\Delta z^{N.S.}_{ \  \parallel}, \ \Delta z^{N.S.}_{\ \perp}  \le \epsilon^3 \frac{1}{(1-a)^2}\ O(1)
\end{equation}
Such a bound is useful for detecting qualitative behavior of the accuracy, however, as is common with asymptotic methods like domain perturbation, the actual error is commonly well under the bound.

To test this, we compute a near-exact flow solution for parallel shear over the family of surfaces (Eq \ref{groove_family}).  Since the velocity is always parallel to the grooves, the problem reduces to solving the Laplace equation for the in-line component $v(x,z)$ under the boundary conditions $v_z(z=\infty)=\tau\equiv1$ and $v(x, \epsilon h(x))=0$.  This can be solved as a superposition of separable solutions
\begin{equation}
v(x,z)=z+\sum_{m=0}^{\infty}A_m \cos(m x) \exp(-m z)
\end{equation}
where the $A_m$ are chosen to satisfy the no-slip boundary condition.  To compute a numerical solution, we solve for the first 1000 terms $A_m$ by enforcing the boundary condition on 1000 equally spaced $x$ values from $0$ to $\pi$.

\begin{figure}
\begin{center}
\epsfig{file=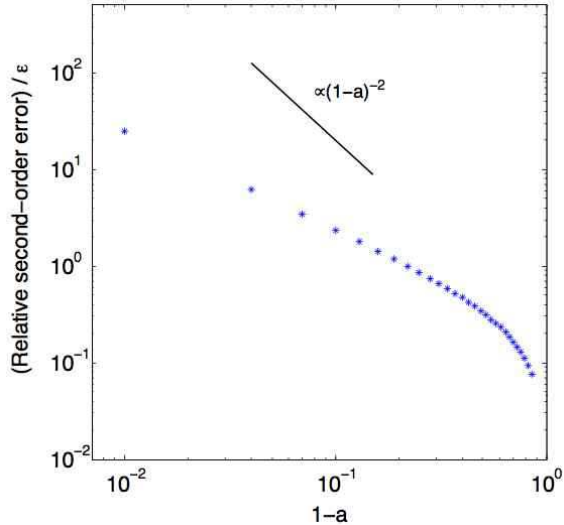,width=3in}
\caption{The true error of the second-order approximation for $z^{N.S.}_{\ \parallel}$ relative to the size of the second-order correction (i.e. $\epsilon^2/8$) vs. the value of $a$. The rate of divergence near $a=1$ is notably lower than our conservative error bound $\propto (1-a)^{-2}$. }\label{error_plot}
\end{center}
\end{figure}

Flows are solved using $a$ values ranging from $0.15$ to $0.99$ in increments of $0.03$.  For each $a$, the flow is computed using seven $\epsilon$ values decreasing from $10^{-1}$ to $10^{-4}$ logarithmically.  Our goal is to determine how accurate Eq \ref{zpar} is compared to the finest term of the approximation, the second-order correction $\epsilon^2/8$.  Hence, for each $a$ and $\epsilon$ pair, we compute the relative second-order error--- the difference between the near-exact numerical solution for $z^{N.S.}_{\ \parallel}$ and the second-order approximation (Eq \ref{zpar}), scaled by the size of the second-order correction term $\epsilon^2/8$.  As expected from the error bound (Eq \ref{groove_bound}), the relative second-order error comes out as being proportional to $\epsilon$ and diverges to $\infty$ as $a\to1$.  However, the proportionality constant is in general much smaller, and diverges more slowly than the bound. The results are plotted in Figure \ref{error_plot}.  From the plot, we see that if $\epsilon=0.1$, for example, then as long as $a$ is less than about $0.8$, the second-order prediction for the no-slip plane is accurate to within one tenth of the size of the second-order correction term.

\section{Optimizing slip properties}\label{opt}

With a direct approximation for the mobility matrix and bounds for the error, we now attempt to answer the question:  What surface shapes minimize/maximize the effective slip? Problems of this type, relating to flow optimization with rigid interfaces, have been looked at primarily in three dimensions, through elegant analyses of optimally porous structures for fluid permeability (see \cite{jung2005}).  Here we study essentially a reduced dimensional problem of optimizing the effective planar slip with respect to a rigid boundary.  We study the case of macroscopic shearing, though in the opposite limit of a thin-channel flow, similar wall-optimization studies have recently been performed \citep{feuillebois2009}.

Let us define
\begin{equation}
\text{Forward mobility}=\frac{\vec{u}^s\cdot(\taub/|\taub|)}{|\taub|}\approx-\frac{\epsilon^2}{|\taub|^2}\left(\taub\cdot\tilde{\vec{M}}(h)\cdot\vec{\taub}\right)
\end{equation}
which measures how much slip occurs in the direction of $\taub$ per unit shear stress.  As is evident from Eq \ref{slipmatrix}, the most forward mobility occurs when the surface is perfectly flat:  If $h(x,y)=const$,  then $\tilde{\vec{M}}=\vec{0}$ and $\vec{u}^s=\vec{0}$ (as expected from a flat no-slip surface). For all other surfaces, $\tilde{\vec{M}}$ is necessarily positive definite and the forward mobility must be negative.  In simple terms, fluctuating surfaces always resist flow more than flat surfaces.

We next investigate the related question of how to maximize/minimize the mobility given that $h(x,y)$ must have some fixed level of heterogeneity.  We quantify the heterogeneity through the variance of $h(x,y)$:
\begin{equation}
\text{Var}(h)=\langle (h(x,y)-\langle h(x,y)\rangle)^2\rangle=\sum_{(m,n)\neq\vec{0}}\left|\hat{h}(m,n)\right|^2. \nonumber
\end{equation}

\subsection{Maximal forward mobility}
Suppose $\epsilon$ is fixed, $L_x=L_y\equiv L$, and $h(x,y)$ is constrained to have a fixed variance $\sigma^2$.  We now derive the surface shape $h(x,y)$ that maximizes the mobility.  This is essentially a Lagrange multiplier problem where the unknowns are the coefficients $|\hat{h}(m,n)|$.  Without loss of generality, presume $\taub$ is aligned with the $x$ direction.  Then our goal is to minimize

\begin{figure}
\begin{center}
(a)\epsfig{file=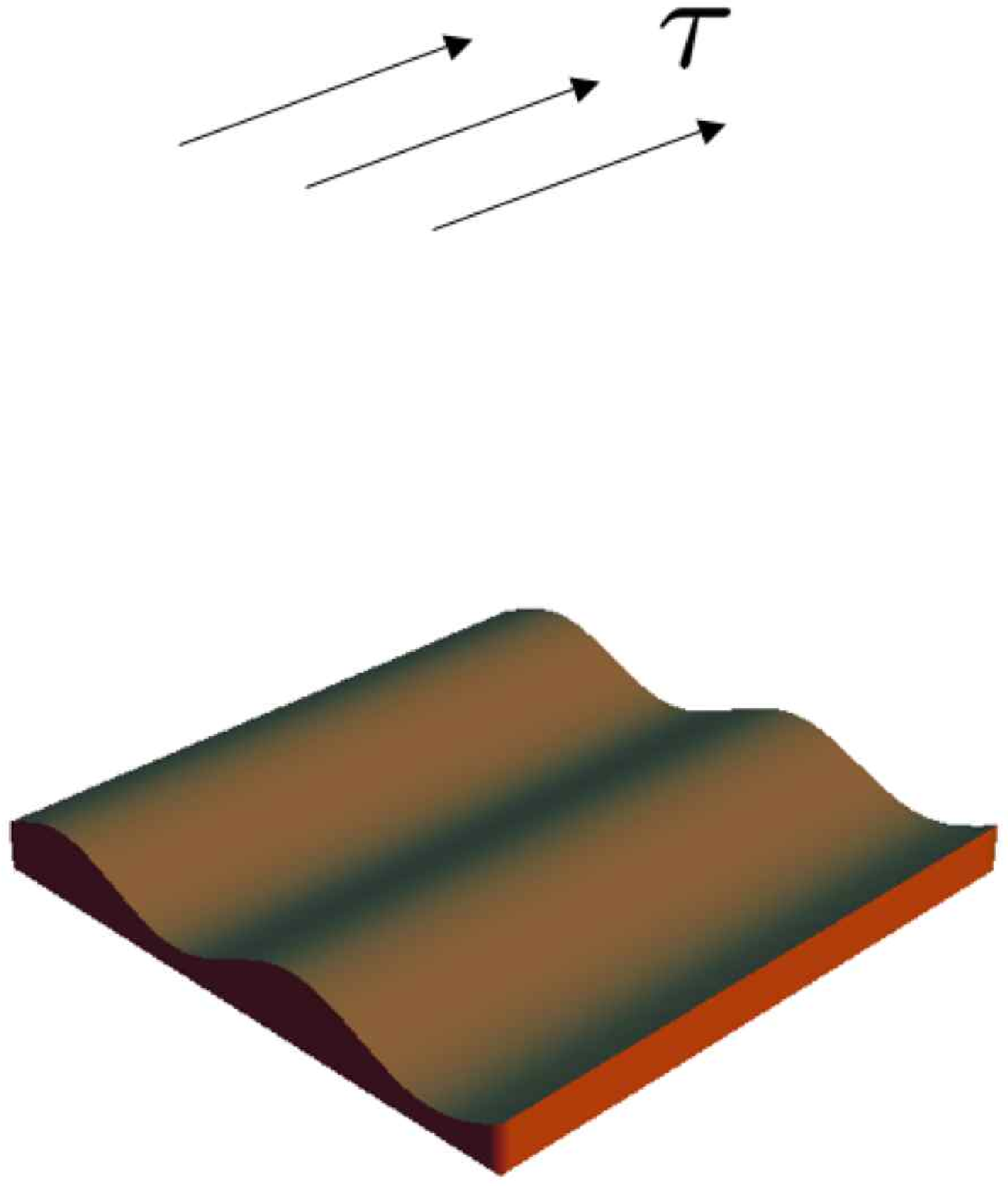,width=2.7in,clip}  \ \ \ \ \ \ \ \ (b)\epsfig{file=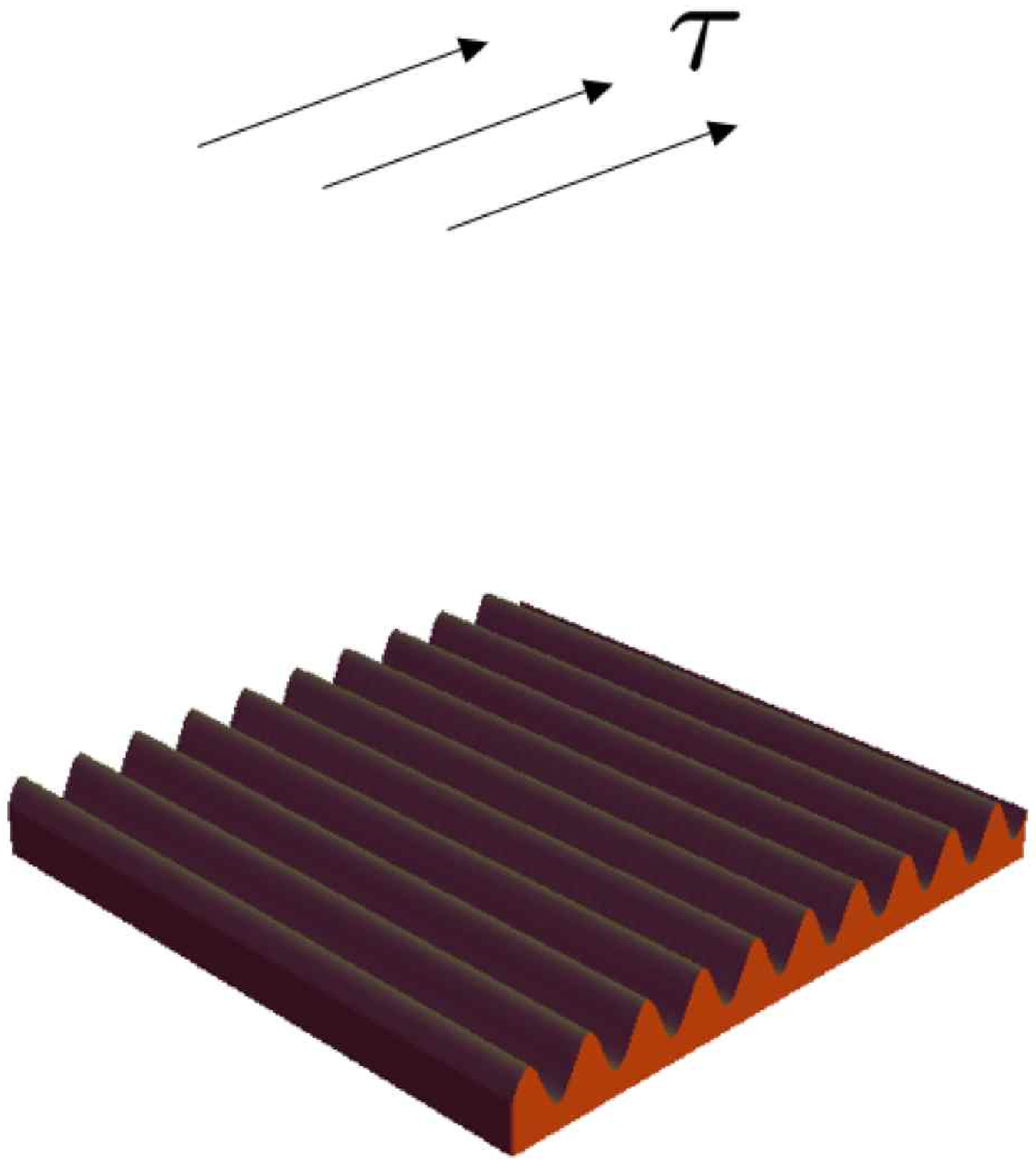,width=2.7in,clip}
\caption{(a)  The periodic surface of fixed variance that restricts flow the least (i.e. maximal forward mobility) has long wavelength sinusoidal grooves oriented along the direction of the stress.  (b)   The surface of fixed variance that restricts flow the most (i.e. minimal forward mobility) has sinusoidal grooves of the shortest possible wavelength oriented across the direction of the stress.}
\end{center}
\end{figure}

\[\hat{\vec{x}}\cdot\tilde{\vec{M}}\cdot\hat{\vec{x}}=\tilde{M}_{11}=\sum_{(m,n)\neq \vec{0}}\frac{2k_m^2+k_n^2}{\sqrt{k_m^2+k_n^2}}\left|\hat{h}(m,n)
\right|^2\]
subject to the constraint
\begin{equation}\label{constraint}
\text{Var}(h)=\sum_{(m,n)\neq\vec{0}}\left|\hat{h}(m,n)\right|^2=\sigma^2.
\end{equation}
The critical points occur only for
\[\nabla_{_{\left|\hat{h}(m,n)\right|}}\ \tilde{M}_{11}(h)=\lambda\ \nabla_{_{\left|\hat{h}(m,n)\right|}}\ \text{Var}(h)\]
for Lagrange multiplier $\lambda$.  Evaluating this form gives
\begin{equation}\label{lagrange}
  2\frac{2k_m^2+k_n^2}{\sqrt{k_m^2+k_n^2}}\left|\hat{h}(m,n)
\right|=2\lambda \left|\hat{h}(m,n)\right| \ \ \  \text{for all} \ \ (m,n)\neq\vec{0}.
\end{equation}
Hence, any surface constituting a critical point of the Lagrange multiplier problem can only contain modes with the same value of $(2k_m^2+k_n^2)/\sqrt{k_m^2+k_n^2}$.  So any critical point containing a certain mode $e^{i(k_{\alpha}x+k_{\beta}y)}$ must, by  Eqs \ref{constraint} and \ref{slipmatrix}, have
\begin{equation}\label{critM}
\tilde{M}_{11}=\frac{2k_{\alpha}^2+k_{\beta}^2}{\sqrt{k_{\alpha}^2+k_{\beta}^2}}\sigma^2.
\end{equation}
Of these critical $\tilde{M}_{11}$ values, the minimum value is $\tilde{M}_{11}=\pi \sigma^2/L$, which occurs only for $(\alpha,\beta)=(0,\pm1)$.  Together with the fact that $\hat{h}(m,n)=\hat{h}(-m,-n)^*$, we obtain
\[\hat{h}(0,1)=\frac{\sigma}{\sqrt{2}}\ e^{i\phi}, \ \ \ \hat{h}(0,-1)=\frac{\sigma}{\sqrt{2}}\ e^{-i\phi},  \ \ \ \text{and} \ \ \ \ \hat{h}(m\neq0,n\neq\pm1)=0\]
for some constant phase $\phi$.  The minimizing surface given our constraints is then
\begin{equation}\label{min}
h(x,y)=\frac{\sigma}{\sqrt{2}}\ e^{i\phi}e^{i \pi y/L}+\frac{\sigma}{\sqrt{2}}\ e^{-i\phi}e^{-i \pi y/L}=\sigma\sqrt{2}\ \cos\left(\frac{\pi y}{L}+\phi\right).
\end{equation}
Thus, at fixed variance, the least resistance to forward flow occurs when the surface is a single-mode sinusoid and the fluid is pushed in the direction \emph{along} the grooves.  Moreover, the ideal wavelength of the surface is the \emph{longest} one allowable for the given periodicity.

\subsection{Minimal forward mobility}
Before determining the surface shape with the least forward mobility, an important caveat must be included:  To wit, we constrict the allowable bandwidth of the surface.  For some positive $K$, consider only scaled surfaces of the form 
\begin{equation}
h(x,y)=\sum_{\substack{{\left\{m,n:  \ 0<\sqrt{k_m^2+k_n^2}<K\right\}}}}\hat{h}(m,n)e^{i(k_mx+k_ny)}
\end{equation}

Capping the bandwidth is the simplest way to avoid surfaces that violate the second-order accuracy of the perturbation expansion.  Naive observation of Eq \ref{critM} would suggest that minimal mobility occurs as $\alpha, \beta\rightarrow \infty$, as this maximizes the critical values of $\tilde{M}_{11}$. But, in view of the error bound Eq \ref{errorwitheps}, we see that such a solution violates the accuracy requirements, as it corresponds to a surface with infinite frequency and hence unbounded slope and curvature.  By considering only surfaces of a finite bandwidth, we guarantee there exists a single, fixed $\epsilon$ so that all critical points of the Lagrange multiplier problem correspond to sufficiently accurate solutions.  This feature is not a major setback in the study of realistic surfaces, where material structure usually possesses some inherent ``grain size'' placing natural limits on the sharpness of corners.

This being said, we reduce to the problem of minimizing the forward mobility over the set of surface shapes $h$ with $\text{Var}(h)=\sigma^2$, and maximal wavenumber $K$.  Let 
\begin{equation}
\Gamma=\lfloor K L/\pi\rfloor.\
\end{equation}
Then by inspection of Eq \ref{critM}, the largest critical $\tilde{M}_{11}$ value is $2\Gamma\pi\sigma^2/L$, which occurs only for $\beta=0$, and $\alpha=\pm\Gamma$.  The corresponding surface is
\begin{equation}
h(x,y)= \frac{\sigma}{\sqrt{2}}\ e^{i\phi}e^{i \Gamma\pi x/L} + \frac{\sigma}{\sqrt{2}}\ e^{-i\phi}e^{-i \Gamma \pi x/L}=\sigma\sqrt{2}\ \cos\left(\frac{\Gamma\pi x}{L}+\phi\right).
\end{equation}
Hence, the surface with the least forward mobility has sinusoidal grooves oriented \emph{perpendicular} to the direction of the shear stress. The wavelength of the grooves should be the \emph{shortest} possible, given the fixed surface periodicity and fixed bandwidth.  In light of the result from the previous subsection, a very natural symmetry arises in the solutions to maximum/minimum mobility.  The overall directionality of the grooves in each case is not surprising --- the transverse and parallel distributions are ubiquitous in the optimization of physical properties for a variety of materials \citep{torquato2002} --- though the groove \emph{shape} is perhaps less obvious.

\section{Random height fluctuations}\label{random}

Based on the prior section, an appropriate follow up is to determine the mobility properties of a ($2L$-periodic) random surface under an identical set of constraints, those being:

\begin{enumerate}
\item  The surface is real, and therefore $\hat{h}(m,n)=\hat{h}(-m,-n)^*$.
\item The surface has $\text{Var}(h)=\sigma^2$.
\item  The bandwidth of $h$ is finite, with maximal wavenumber $K$.
\end{enumerate}

Let $A$ be the set of integer pairs $(m,n)$ with the property that $\sqrt{k_m^2+k_n^2}<K$, $m\ge0$ when $n>0$, and $m>0$ whenever $n\le0$.   Under these definitions, any surface obeying constraints (1) and (3) above must be of the form
\begin{equation}
h(x,y)=\hat{h}(0,0)+\sum_{(m,n)\in A}\left(\hat{h}(m,n)e^{i(k_mx+k_ny)}+\hat{h}(m,n)^*\ e^{-i(k_mx+k_ny)}\right)
\end{equation}
and the variance constraint (2) is
\begin{equation}
\sum_{(m,n)\in A}2|\hat{h}(m,n)|^2=\sigma^2.
\end{equation}

Observing the slip matrix formula Eq \ref{sliprelation}, the mobilities depend only on the magnitude of the Fourier coefficients. This suggests a simple, broad probability distribution for sampling the surfaces as shall be described next.   First, index the members of $A$ by $\vec{p}_1$ through $\vec{p}_{|A|}$ and define $\gamma_{j}\equiv2|\hat{h}(\vec{p}_j)|^2$.  The region of space in the variables $\{\gamma_1,...,\gamma_{|A|}\}$ that fulfills the constraint on the variance is expressed by

\begin{figure}
\begin{center}
\epsfig{file=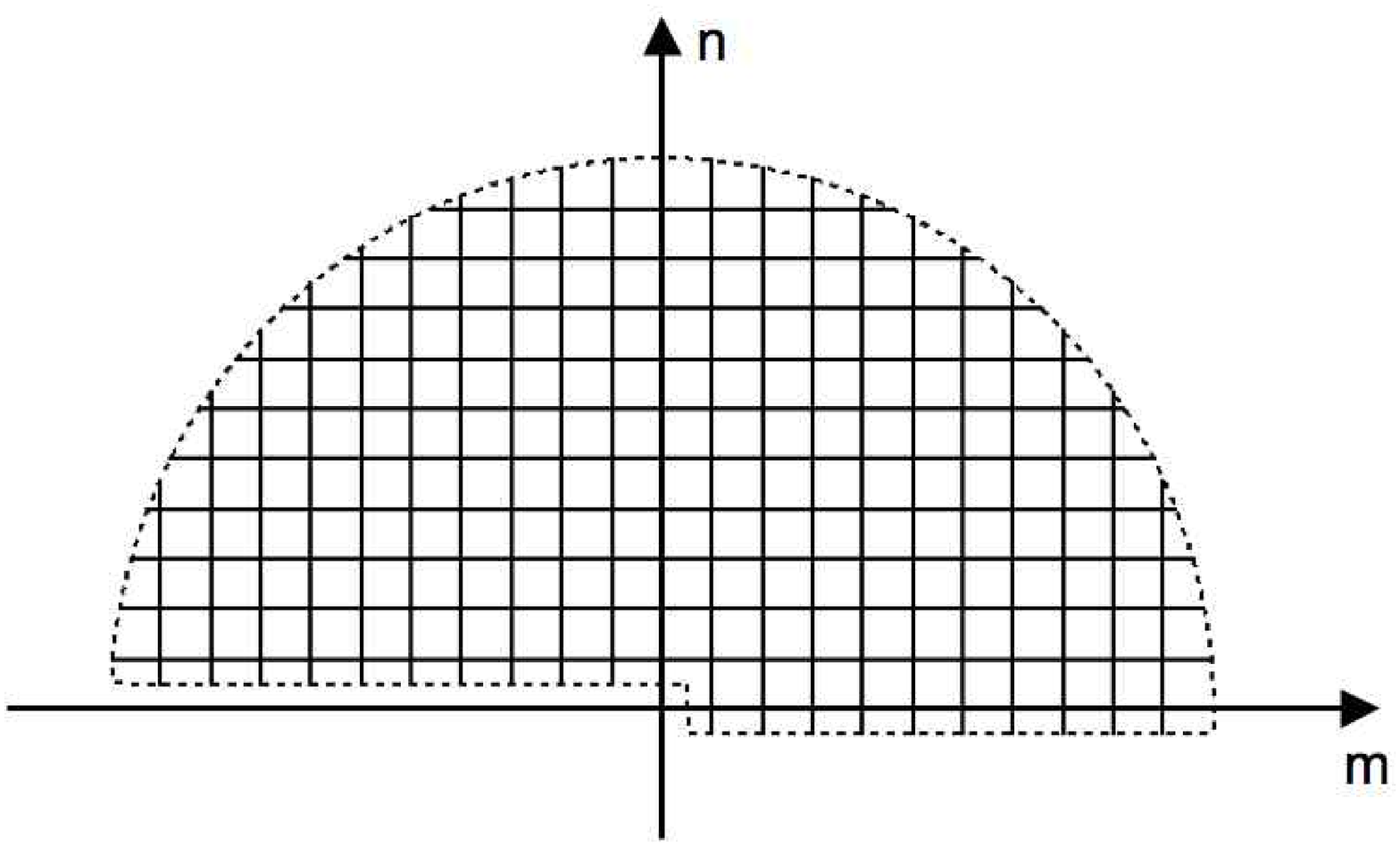, width=4in}
\caption{The integer grid points within an upward semicircle, excluding the non-positive $m$-axis, compose the set $A$.  Up to an additive constant, any periodic surface of finite bandwidth can be uniquely described by determining $\hat{h}(m,n)$ on $A$.}
\end{center}
\end{figure}
\begin{subeqnarray}
&& 0\le\gamma_1 \le \sigma^2
\\
&& 0\le\gamma_2 \le \sigma^2-\gamma_1
\\
&& 0\le\gamma_3 \le \sigma^2-\gamma_1-\gamma_2
\\
 && \ \ \ \ \ \ \ \vdots \nonumber
\\
 && 0\le\gamma_{|A|-1} \le \sigma^2-\sum_{j=1}^{|A|-2}\gamma_j
\\
&&  \gamma_{|A|} =\sigma^2-\sum_{j=1}^{|A|-1}\gamma_j \ .
\end{subeqnarray}

Let us denote by $\Omega$ the higher dimensional tetrahedron described by the equalities above on $\gamma_1, \cdots, \gamma_{|A|-1}$.  Each surface $h$ obeying our constraints is equivalently one point in the region $\Omega$. So, one probability distribution we can write is \\

\begin{center}
\begin{tabular}{cccl}
\Bigg\{\parbox{6cm}{Prob of $h\in\Omega$ with $\gamma$'s between $\gamma_j$ and $\gamma_j+d\gamma_j$ for each $1<j<|A|-1$}\Bigg\}
&=
& \parbox{4.7cm}{
\begin{equation*}
 \prod_{j=1}^{|A|-1}d\gamma_j \left/ \left(\int_\Omega \prod_{j=1}^{|A|-1}d\gamma_j\right)\right.
\end{equation*}}
&
\parbox{2cm}{\begin{equation}
\end{equation}}
\end{tabular}
\end{center}

Under this distribution, the probability of choosing some surface within a subset of $\Omega$ occurs with probability proportional to the volume of that subset.  Using an overbar to denote the weighted average by this distribution, basic integration gives the identity

\begin{equation}
\overline{\sum_{j=1}^{|A|}f(\vec{p}_j)\gamma_j}\ =\ \frac{\sigma^2}{|A|}\sum_{j=1}^{|A|}f(\vec{p}_j)
\end{equation}
for any function $f$.  Applying this to Eq \ref{slipmatrix}, one finds
\begin{equation}\label{matrix}
\overline{\tilde{\vec{M}}}=\frac{\sigma^2}{|A|}\begin{pmatrix}
\displaystyle\sum_{(m,n)\in A}\frac{2k_m^2+k_n^2}{\sqrt{k_m^2+k_n^2}} 
& \displaystyle\sum_{(m,n)\in A}\frac{k_mk_n}{\sqrt{k_m^2+k_n^2}}
\\
\displaystyle \sum_{(m,n)\in A}\frac{k_mk_n}{\sqrt{k_m^2+k_n^2}} 
& \displaystyle\sum_{(m,n)\in A}\frac{k_m^2+2k_n^2}{\sqrt{k_m^2+k_n^2}}
\end{pmatrix}.
\end{equation}
Next, each of these sums over $A$ must be computed.  Each sum in the matrix above can be approximated by the semicircular integral
\[\sum_{(m,n)\in A} f(m,n)\cong \int_{r=0}^{KL/\pi}\int_{\theta=0}^{\pi}f(r\cos\theta,r\sin\theta)\ r dr d\theta \]
 due to symmetry along the $m$-axis and the fact the $f(0,0)=0$ for each summand.  Then we may write
 \begin{equation}
 \overline{\tilde{\vec{M}}}\cong\frac{\sigma^2}{|A|}\begin{pmatrix}
\frac{K^3L^2}{2\pi}& 0
\\
0 & \frac{K^3L^2}{2\pi}
\end{pmatrix}=\sigma^2\frac{K^3L^2}{2\pi|A|}\vec{1}.
\end{equation}
where $\vec{1}$ is the $2\times2$ identity tensor. Lastly, the number of elements in $A$ can be computed with a semicircular area approximation (excluding the missing point at the origin):
\[|A|\cong\frac{1}{2}\pi(KL/\pi)^2-1\]
giving the final result
\begin{equation}
 \overline{\tilde{\vec{M}}}\cong \sigma^2\frac{K^3L^2}{ K^2L^2-2\pi }\vec{1} \cong \sigma^2 K \vec{1}.
\end{equation}
where the last approximation holds for $KL$ large enough.

As expected, a random surface of fixed variance and bandwidth has isotropic slip properties. Moreover, the forward mobility of a random surface is approximately $-\epsilon^2\sigma^2K$ which fits nicely into the ordering we find for the most and least optimal surfaces of fixed variance as discussed in section \ref{opt}:
\[\text{Forward mobility:} \ \ \ \ \ \ \ \ \ \underbrace{-\epsilon^2\sigma^2\frac{\pi}{L}}_{\text{Most}}<\underbrace{-\epsilon^2\sigma^2 K}_{\text{Random}}< \underbrace{-2\epsilon^2\sigma^2 \frac{\Gamma\pi}{L}}_{\text{Least}}.\]
Or, perhaps the ordering is easier to see in terms of the relative sizes of these three quantities, which is closely approximated by
\[\text{Most}:\text{Random}:\text{Least}\ \cong\ -\left(1:\Gamma:2\Gamma\right)\]
But we reiterate that the result for mean mobility over a random surface depends entirely on the choice of the probability distribution.  We have selected a particular one, which is simple enough to perform the necessary calculations, and which appears relatively unbiased toward any particular subset of surfaces.

\section{Surfaces with fluctuating scalar Navier slip}\label{navier}

Up to this point, the focus has been on surface patterns due solely to height fluctuations.  Now, let us expand the analysis to surfaces that have fluctuating scalar slip properties, such as composite surfaces where each material has different hydrophobicity properties.

Such surfaces can be described with a scalar Navier slip boundary condition, which relates the fluid slip along the surface to the fluid shear-rate at the surface
\begin{equation}
\vec{u}^s=b(x,y) \ \deriv{\vec{u}}{n}. 
\end{equation}
A no-slip surface corresponds to a Navier slip coefficient of $b=0$, whereas a traction-free surface corresponds to $b=\infty$. In this section, we analyze the macroscopic flow properties of surfaces with periodic slip coefficient $\epsilon b(x,y)$ measuring small variations from no-slip,  as well as possible height fluctuations $\epsilon h(x,y)$.  Applying the Navier slip condition along the surface $\epsilon h(x,y)$ gives the boundary condition
\begin{equation}
 \vec{u}(x,y,\epsilon h(x,y))= \epsilon b(x,y) \  \frac{(-\epsilon h_x, \ -\epsilon h_y, \ 1)}{\left| (-\epsilon h_x, \ -\epsilon h_y, \ 1)\right| } \cdot \nabla \vec{u}\Big{|}_{z=\epsilon h(x,y)} 
 \end{equation}

Adopting the perturbation series representations of $\vec{u}$ and $p$ from section \ref{pert_series}, one expands in $\epsilon$ to yield the following term-by-term conditions for the flow at $z=0$:

\begin{align}
\vec{u}_0(x,y,0)&=\vec{0}
\\
\vec{u}_1(x,y,0)&=(b(x,y)-h(x,y))\ \left.\deriv{\vec{u}_0}{z}\right|_{z=0}
\\
\vec{u}_2(x,y,0)&=-h(x,y)\left.\deriv{\vec{u}_1}{z}\right|_{z=0}-\frac{h(x,y)^2}{2}\left.\deriv{^2\vec{u}_0}{z^2}\right|_{z=0}
\\
&\hspace{1cm}+b(x,y)\left(h(x,y)\deriv{^2\vec{u}_0}{z^2}  +\left(-h_x\deriv{\vec{u}_0}{x}\Big|_{z=0}, \  -h_y\deriv{\vec{u}_0}{y}\Big|_{z=0},\ 1\right) \cdot\nabla\vec{u}_1\Big|_{z=0} \right)
\end{align}
The $z=\infty$ conditions are the same as those in section \ref{pert_series}.  The term $\vec{u}_0$ is solved, as before, by
\begin{equation}\vec{u}_0(x,y,z)=\taub z.\end{equation}
Substituting this result into the above boundary conditions for the other orders, one finds
\begin{align}
\vec{u}_1(x,y,0)&=(b(x,y)-h(x,y))\ \taub
\\
\vec{u}_2(x,y,0)&=(b(x,y)-h(x,y))\ \left.\deriv{\vec{u}_1}{z}\right|_{z=0}.
\end{align}
Comparing to sections \ref{first} and \ref{second}, it is now evident that these boundary conditions are the same as the previous, except with the switch $h(x,y)\rightarrow h(x,y)-b(x,y)$.  Hence, the solution up to second-order can be inferred directly by modifying the results of the previous sections.

Keeping with the convention that $z=0$ at the bottom of the surface, we apply the switch and obtain the result
\begin{equation}\vec{u}(z\rightarrow\infty)=\taub z-\epsilon\langle h(x,y)-b(x,y)\rangle\taub -\epsilon^2\tilde{\vec{M}}(h-b)\cdot\taub+O(\epsilon^3)\end{equation}
Writing, as before, in terms of the variable $\tilde{z}=z-\epsilon\langle h(x,y)\rangle$, one obtains the slip relation
\begin{equation}\label{heightslip}
\vec{u}^s=\epsilon\langle b(x,y) \rangle\taub-\epsilon^2 \tilde{\vec{M}}(h-b)\cdot\taub+O(\epsilon^3).
\end{equation}
Up to second-order,  the effective slip of a surface with small scalar slip and height fluctuations is the same as that of a no-slip surface except with $h\rightarrow h-b$ in the argument of $\tilde{\vec{M}}$, plus an isotropic first-order term due to the average of $b$.

\subsection{A simple optimization for surfaces with height and Navier slip fluctuations}

Consider a surface with height and slip fluctuations of the same periodicity.  The forward mobility, as computed from Eq \ref{heightslip}, is
\begin{equation}
\frac{\vec{u}^s\cdot\taub}{|\taub|^2}\approx\epsilon \langle b(x,y) \rangle -\epsilon^2\frac{\taub\cdot\tilde{\vec{M}}(h-b)\cdot\taub}{|\taub|^2}.\end{equation}
Holding the average of $b(x,y)$ fixed, the mobility can be maximized by observing the second-order term above.  As per the definition of $\tilde{\vec{M}}$ (Eq \ref{slipmatrix}), the mobility matrix is positive definite if $b(x,y)-h(x,y)\neq const$, and $\vec{0}$ otherwise. Consequently, the mobility is maximized when $b(x,y)-h(x,y)=const$. For any given $h(x,y)$, the optimal choice of $b(x,y)$ is
\begin{equation}
b(x,y)-\langle b(x,y) \rangle=h(x,y)-\langle h(x,y) \rangle
\end{equation}
which gives a mobility of $\epsilon\langle b(x,y)\rangle$. Physically speaking, this means the forward slip is maximized when the peaks of the surface $h$ are more hydrophobic, and the valleys of the surface are more hydrophilic.

\section{Conclusion}

This work has derived a second-order accurate formula (Eqs \ref{sliprelation} and \ref{slipmatrix}) describing an effective local boundary condition for shear flows over small-fluctuation periodic surfaces.  The formula represents a tensorial mobility law, and is easily extended to include surfaces of both non-uniform hydrophobicity and height changes  (Eq \ref{heightslip}). We have gone to great lengths to quantify the error of the approximation, so as to provide firm guidelines for its appropriate usage.  Within these guidelines, the formula was optimized in a Lagrange multiplier framework to derive the no-slip surfaces of fixed variance that maximize/minimize the forward mobility.
These extremal cases were favorably compared with the mobility of a random, fixed-variance surface as computed by summing over some unbiased distribution.  We have also performed a simple optimization that instructs the optimal coupling between height and scalar slip fluctuations when both surface effects can take place.

In the future, we hope to augment our analysis to include the possibility of a surface charge profile.  This addition would be useful in electro-kinetic applications to help understand and predict electro-osmotic fluid transport.  We also continue from a mathematical perspective to work on enhanced solution methods that may improve the breadth of applicability of the mobility formula.

\appendix{

\section{Appendix: General solution to the Stokes Equations}\label{stokes}

In this appendix, a general solution to the Stokes equations is given for use in solving the first- and second-order terms in the perturbation expansion displayed in Section \ref{pert_series}.  It is expressed as a Fourier series in the horizontal dimensions with three undetermined coefficient sets:

\begin{align*}
u(x,& y,z)=\alpha+\sum_{(m,n)\neq\vec{0}}e^{i(k_m x+k_ny)}\frac{e^{-z\sqrt{k_m^2+k_n^2}}}{3\sqrt{k_m^2+k_n^2}}\left[-ik_m\left(1+2z\sqrt{k_m^2+k_n^2}\right)A(m,n) \right.
\\
&\left.+\frac{3k_n^2\sqrt{k_m^2+k_n^2}-2k_m^4z+2k_m^2(\sqrt{k_m^2+k_n^2}-k_n^2z)}{k_m^2+k_n^2}B(m,n)
- \frac{k_mk_n(\sqrt{k_m^2+k_n^2}-2z(k_m^2+k_n^2))}{k_m^2+k_n^2}C(m,n)\right]
\\
\\
v(x,& y,z)=\beta+\sum_{(m,n)\neq\vec{0}}e^{i(k_m x+k_ny)}\frac{e^{-z\sqrt{k_m^2+k_n^2}}}{3\sqrt{k_m^2+k_n^2}}\left[-ik_n\left(1+2z\sqrt{k_m^2+k_n^2}\right)A(m,n) \right.
\\
&\left.-\frac{k_mk_n(\sqrt{k_m^2+k_n^2}-2z(k_m^2+k_n^2))}{k_m^2+k_n^2}B(m,n)+
\frac{3k_m^2\sqrt{k_m^2+k_n^2}-2k_n^4z+2k_n^2(\sqrt{k_m^2+k_n^2}-k_m^2z)}{k_m^2+k_n^2}C(m,n)\right]
\\
\\
w(x,& y,z)=\gamma+\sum_{(m,n)\neq\vec{0}}e^{i(k_m x+k_ny)}\frac{e^{-z\sqrt{k_m^2+k_n^2}}}{3}\left[\left(3+2z\sqrt{k_m^2+k_n^2}\right)A(m,n)-2ik_mz B(x,y)-2ik_nzC(m,n)\right]
\\
\\
p(x,& y,z)=\delta+\sum_{(m,n)\neq\vec{0}}e^{i(k_m x+k_ny)}\frac{e^{-z\sqrt{k_m^2+k_n^2}}}{3}\left[4\sqrt{k_m^2+k_n^2}A(m,n)-4ik_mB(m,n)-4ik_nC(m,n)\right]
\end{align*}

This system can be inverted to uniquely determine the constants $\alpha,\ \beta,\ \gamma$ and the coefficient sets $A(m,n),\ B(m,n),\ C(m,n)$ in terms of the chosen lower boundary condition on the velocity, $\vec{u}(x,y,0)$. The general solution automatically upholds the known boundary conditions for the first- and second-order perturbation terms:
\[ \vec{u}(x,y,z)=\vec{u}(x+L_x,y+L_y,z) \ \  \text{,} \ \ \ \  p(x,y,z)=p(x+L_x,y+L_y,z) \ \ \text{,} \ \ \ \ \left.\deriv{\vec{u}}{z}\right|_{z= \infty}=\vec{0}.\]

\section{Appendix:  Determining error bounds}\label{derivation}


This appendix derives the error bound displayed as Eq \ref{errorwitheps} in Section \ref{bounderror}.  To begin, define the following norm on scalar, vector, or tensor fields $\vec{f}(x,y)$ over a 2D periodic domain:
\begin{equation}
\left\|\vec{f}(x,y)\right\|_{x,y}=\sqrt{\frac{1}{4L_xL_y}\int_{-L_x}^{L_x}\int_{-L_y}^{L_y}|\vec{f}(x,y)|^2 \ dxdy}
\end{equation}
We reserve $|\cdot|$ for the absolute value (if applied to a scalar) or Euclidean norm (if applied to a vector or matrix). The above is directly proportional to the compact support $L_2$ norm on functions.  The choice of this norm greatly simplifies the analysis since it it connects directly to an inner product, and consequently an array of useful theorems.  

The exact flow solution in our problem satisfies the no-slip condition along the surface, i.e. $\vec{u}(x,y,\epsilon h(x,y))=\vec{0}$. To measure how much our second-order flow approximation differs from the true solution along the surface, we use the norm to measure the surface error:
\begin{align}
\text{Surface Error}&=\left\| \vec{u}_0(x,y,\epsilon h(x,y))+\epsilon\vec{u}_1(x,y,\epsilon h(x,y))+\epsilon^2\vec{u}_2(x,y,\epsilon h(x,y))\right\|_{x,y}
\end{align}
From Taylor's Remainder Theorem, we know
\begin{align*}
\vec{u}_1(x,y,\epsilon h)&=\vec{u}_1(x,y,0)+\epsilon h\left.\deriv{\vec{u}_1}{z}\right|_{z=0}+\int_0^{\epsilon h(x,y)}(\epsilon h(x,y)-z)\deriv{^2\vec{u_1}}{z^2}\ dz
\\
\vec{u}_2(x,y,\epsilon h)&=\vec{u}_2(x,y,0)+\int_0^{\epsilon h(x,y)}\deriv{\vec{u_2}}{z}\ dz.
\end{align*}
Inserting these forms into the above, and using the order-by-order boundary conditions to cancel terms, one obtains
\begin{align}
\text{Surface Error}&=\left\| \epsilon\left(\int_0^{\epsilon h(x,y)}(\epsilon h(x,y)-z)\deriv{^2\vec{u_1}}{z^2}\ dz\right)+\epsilon^2\left(\int_0^{\epsilon h(x,y)}\deriv{\vec{u_2}}{z}\ dz\right)\right\|_{x,y} \nonumber
\\
&\le \epsilon \underbrace{\left\| \int_0^{\epsilon h(x,y)}(\epsilon h(x,y)-z)\deriv{^2\vec{u_1}}{z^2}\ dz  \right\|_{x,y} }_{\equiv R^{(1)}} +  \epsilon^2 \underbrace{\left\| \int_0^{\epsilon h(x,y)}\deriv{\vec{u_2}}{z}\ dz \right\|_{x,y}}_{\equiv R^{(2)}} \label{rr}
\end{align}
by the triangle ininequality.

To place an upper bound on the error, we must calculate tight bounds on each of these two integrals given any surface $h(x,y)$.  This is a highly nontrivial task.  As is derived in Appendix \ref{r1r2}, with judicious use of Parseval's Theorem and the Cauchy-Schwartz inequality, we can show that
\begin{align}
&R^{(1)}\equiv\left\|\int_0^{\epsilon h(x,y)}(\epsilon h(x,y)-z)\deriv{^2\vec{u_1}}{z^2}\ dz\right\|_{x,y}\le \epsilon^2 h_M^2 K_1|\taub|\ \left\|\nabla\nabla h(x,y)\right\|_{x,y} \label{r1}
\\
&R^{(2)}\equiv\left\|\int_0^{\epsilon h(x,y)}\deriv{\vec{u_2}}{z}\ dz\right\|_{x,y}\le \epsilon h_M K_2|\taub|\left(|\nabla h|_M\ \|\nabla h(x,y)\|_{x,y}+h_M\ \left\|\nabla\nabla h(x,y)\right\|_{x,y}    \right) \label{r2}
\end{align}
where $K_1$ and $K_2$ are order one, known, dimensionless constants independent of the choice of $\epsilon$, $\taub$, or $h(x,y)$, and for any function $f(x,y)$ we adopt the notation  

\begin{equation}
f_M\equiv\max_{x,y} f(x,y).
\end{equation}  
The bounds on $R^{(1)}$ and $R^{(2)}$ applied to Eq \ref{rr} give
\begin{align}
\text{Surface Error} &\le \kappa \ \epsilon^3 h_M |\taub| \big(h_M\ \|\nabla\nabla h(x,y)\|_{x,y}+|\nabla h|_M\ \|\nabla h(x,y)\|_{x,y}\big) \nonumber
\\
&\le \kappa \ \epsilon^3 h_M |\taub| \big(h_M\ \left|\nabla\nabla h\right|_M+|\nabla h|_M^2\big) \equiv\text{Surface Error Bound} \label{surferror}
\end{align}
where we define $\kappa \ \equiv K_1+K_2$, which is another known order-one constant.

Next, we make the connection between the surface error and the error in the predicted slip. The exact flow solution must have the velocity go to zero along the surface. The second-order approximation that we have found has some fallacious extra velocity along the surface that arises due to truncation error, and the surface error is a space-average measurement of this erroneous surface speed.  If we let $\vec{u}(x,y,z)$ represent the true solution, $\vec{u}^{app}(x,y,z)=\vec{u}_0(x,y,z)+\epsilon\vec{u}_1(x,y,z)+\epsilon^2\vec{u}_2(x,y,z)$ be our second-order approximation, and  $\Deltab(x,y,z)\equiv\vec{u}(x,y,z)-\vec{u}^{app}(x,y,z)$, it follows that
\begin{equation}\label{delta}
\left\|\Deltab(x,y,\epsilon h(x,y))\right\|_{x,y}\le \text{Surface Error Bound}
\end{equation}

Since both $\vec{u}(x,y,z)$ and $\vec{u}^{app}(x,y,z)$ obey the Stokes equations exactly, the difference $\Deltab(x,y,z)$ must also obey the Stokes equations.  Moreover, since both $\vec{u}$ and $\vec{u}^{app}$ satisfy the same traction boundary condition as $z\rightarrow\infty$, then the difference flow $\Deltab(x,y,z)$ must asymptote to a constant uniform flow (zero shear traction) for large $z$.  The speed in the uniform flow region is precisely the error in the approximation for the effective slip.  Our goal is hence reduced to determining the maximum possible flow speed that can occur at large $z$ in a Stokes flow $\Deltab(x,y,z)$ that obeys Eq \ref{delta} and zero shear tractions at $z=\infty$. 

While the total surface error is described by (\ref{surferror}), the distribution of this extra velocity over the surface is undetermined, providing the only degree of freedom in this maximization.  We state without proof that the maximal large-$z$ uniform flow arises for
\begin{equation}
\Deltab(x,y,z)=(\text{Surface Error Bound})\ \vec{e}
\end{equation}
where $\vec{e}$ is any horizontal unit vector.  In other words, $\Deltab$ has a large-$z$ uniform flow of maximal speed when the extra surface velocity represented by the surface error is equally distributed and uniformly directed along the surface, so that $\Deltab(x,y,z)$ is globally uniform. This choice of $\Deltab$ saturates Eq \ref{delta}.

Recalling the definition in  (\ref{surferror}), the above selection of $\Deltab(x,y,z)$  implies the error in the second-order slip approximation is subject to
\begin{equation}
\left|\vec{u}^s-\left(-\epsilon^2\tilde{\vec{M}}(h)\cdot\taub\right)\right|\le \epsilon^3 \kappa \ h_M |\taub| \big(h_M\ \left|\nabla\nabla h\right|_M+|\nabla h|_M^2\big) \nonumber
 \end{equation}
which is our desired result.

\section{Appendix: Bounding the integrals $R^{(1)}$ and $R^{(2)}$}\label{r1r2}
The purpose of this appendix is to rigorously prove the bounds on the error quantities $R^{(1)}$ and $R^{(2)}$ shown as Eqs \ref{r1} and \ref{r2} in Appendix \ref{derivation}.
\subsection{The error $R^{(1)}$}

We begin with the bound on $R^{(1)}$.  First, we observe by the Cauchy-Schwartz inequality that
\begin{align*}
\left|\int_0^{\epsilon h(x,y)}(\epsilon h(x,y)-z)\deriv{^2\vec{u}_1}{z^2}\ dz \right| & \le  \sqrt{\int_0^{\epsilon h(x,y)}\left|(\epsilon h(x,y)-z)\deriv{^2 \vec{u}_1}{z^2}\right|^2 dz} \ \sqrt{\int_0^{\epsilon h(x,y)}1^2\ dz}
\\
& \le \sqrt{\epsilon h(x,y)}\sqrt{\int_0^{\epsilon h(x,y)}(\epsilon h(x,y)-z)^2 \left| \deriv{^2\vec{u}_1}{z^2}\right|^2 dz} 
\\
& \le  \sqrt{\epsilon h_M}\sqrt{\int_0^{\epsilon h_M}(\epsilon h_M-z)^2 \left| \deriv{^2\vec{u}_1}{z^2}\right|^2 dz} 
\end{align*}
We insert this result into the definition of $R^{(1)}$, and expand the function norm:
\begin{align}
R^{(1)} &\le \left\| \sqrt{\epsilon h_M}\sqrt{\int_0^{\epsilon h_M}(\epsilon h_M-z)^2 \left|\deriv{^2 \vec{u}_1}{z^2}\right|^2 dz} \right\|_{x,y}\nonumber 
\\
&= \sqrt{\frac{1}{4L_xL_y}\int_{-L_x}^{L_x}\int_{-L_y}^{L_y}\epsilon h_M \left(\int_0^{\epsilon h_M}(\epsilon h_M-z)^2\left|\deriv{^2 \vec{u}_1}{z^2}\right|^2 dz\right)dxdy} \nonumber
\\
&=\sqrt{\epsilon h_M\int_0^{\epsilon h_M}(\epsilon h_M-z)^2\left\|\deriv{^2 \vec{u}_1}{z^2}  \right\|_{x,y}^2 dz}. \label{integ}
\end{align}
The final line follows from rearranging the order of integration. In order to apply the next step,  we must first compute $|\partial^2 \vec{u}_1/\partial z^2 |^2$, which is straightforward because $\vec{u}_1(x,y,z)$ is completely known.  The result is lengthy, but can be written somewhat compactly as
\begin{align*}
&\left| \deriv{^2 \vec{u}_1}{z^2}\right|^2 = \left( \deriv{^2 u_1}{z^2}\right)^2+\left( \deriv{^2 v_1}{z^2}\right)^2+\left( \deriv{^2 w_1}{z^2}\right)^2
\\
&\ \ \ \ \ =\sum_{j=\{1,2,3\}} \left( \sum_{(m,n)\neq\vec{0}}\ \taub\cdot \left(\vec{q}_{m,n}\cdot \vec{A}_{(j)} \cdot\vec{p}_{m,n}\ \ , \ \ \vec{q}_{m,n}\cdot \vec{B}_{(j)}\cdot \vec{p}_{m,n}\right)  \ e^{-z\sqrt{k_m^2+k_n^2}}\hat{h}(m,n)e^{i(k_m x+k_ny)}\right)^2
\end{align*}
where for each choice of $m$ and $n$, $\vec{q}_{m,n}$ and $\vec{p}_{m,n}$ are the vectors defined by 
\begin{equation}\label{pq}
\vec{q}_{m,n}=(1,z\sqrt{k_m^2+k_n^2},zk_m,zk_n) \ \ \ \text{and} \ \ \ \vec{p}_{m,n}=(k_m^2,k_n^2,\sqrt{2}k_mk_n, k_m\sqrt{k_m^2+k_n^2},k_n\sqrt{k_m^2+k_n^2})
\end{equation}
and for each $j$, $\vec{A}_{(j)}$ and $\vec{B}_{(j)}$ are constant, dimensionless,  $4\times 5$ matrices of order one in size. 

Now for the crucial step: Apply Parseval's Theorem to convert the function norm to a discrete sum,
\begin{align*}
&\left\|\deriv{^2 \vec{u}_1}{z^2}  \right\|_{x,y}^2 =\frac{1}{4L_xL_y}\int_{-L_x}^{L_x}\int_{-L_y}^{L_y}\left| \deriv{^2 \vec{u}_1}{z^2}\right|^2 \ dxdy
\\
& \ \ \ \ \ \ =\sum_{j=\{1,2,3\}}\left[ \frac{1}{4L_xL_y}\int_{-L_x}^{L_x}\int_{-L_y}^{L_y}\left(\sum_{(m,n)\neq\vec{0}}\  \taub\cdot \left(\vec{q}_{m,n}\cdot \vec{A}_{(j)} \cdot\vec{p}_{m,n}\ \ , \ \ \vec{q}_{m,n}\cdot \vec{B}_{(j)}\cdot \vec{p}_{m,n}\right)  \right.\right.
\\
&\hspace{10.5cm} \left.\left.e^{-z\sqrt{k_m^2+_n^2}}\hat{h}(m,n)e^{i(k_m x+k_ny)}\right)^2\right]dxdy
\\
&\ \ \ \ \ \ =\sum_{j=\{1,2,3\}} \sum_{(m,n)\neq\vec{0}}\left| \taub\cdot \left(\vec{q}_{m,n}\cdot \vec{A}_{(j)} \cdot\vec{p}_{m,n}\ \ , \ \ \vec{q}_{m,n}\cdot \vec{B}_{(j)}\cdot \vec{p}_{m,n}\right)  \ e^{-z\sqrt{k_m^2+k_n^2}}\hat{h}(m,n)\right|^2
\\
& \ \ \ \ \ \ \le \sum_{j=\{1,2,3\}} \sum_{(m,n)\neq\vec{0}}|\taub|^2 |\vec{q}_{m,n}|^2\left(\left|\vec{A}_{(j)}\right|^2 +\left|\vec{B}_{(j)}\right|^2\right)|\vec{p}_{m,n}|^2\ e^{-2z\sqrt{k_m^2+k_n^2}}\left|\hat{h}(m,n)\right|^2
\end{align*}
The last line follows from the Cauchy-Schwartz inequality.  To express the result more simply, define 
\[\vec{\alpha}=\sum_{j=\{1,2,3\}}|\vec{A}_{(j)}|^2+|\vec{B}_{(j)}|^2\]
Then, expanding the norms on the vectors $\vec{p}$ and $\vec{q}$, we have
\begin{equation}
\left\|\deriv{^2 \vec{u}_1}{z^2}  \right\|_{x,y}^2 \le \ \alpha |\taub|^2 \sum_{(m,n)\neq\vec{0}}(1+z^2(2k_m^2+2k_n^2))(2k_m^4+2k_n^4+4k_m^2k_n^2)\ e^{-2z\sqrt{k_m^2+k_n^2}}\left|\hat{h}(m,n)\right|^2
\end{equation}

\noindent With a bound on the integrand in (\ref{integ}) now established, we move on to evaluate the integral:
\begin{align}
&\int_0^{\epsilon h_M}(\epsilon h_M-z)^2 \left\|\deriv{^2\vec{u}_1}{z^2}  \right\|_{x,y}^2 dz \nonumber
\\
& \le\int_0^{\epsilon h_M}\left[(\epsilon h_M -z)^2 \alpha |\taub|^2 \sum_{(m,n)\neq\vec{0}}(1+z^2(2k_m^2+2k_n^2))(2k_m^4+2k_n^4+4k_m^2k_n^2)\ e^{-2z\sqrt{k_m^2+k_n^2}}\left|\hat{h}(m,n)\right|^2\right] dz \nonumber
\\
&=2\alpha|\taub|^2 \sum_{(m,n)\neq\vec{0}}\ (k_m^4+k_n^4+2k_m^2k_n^2) \left|\hat{h}(m,n)\right|^2 \left[ \int_0^{\epsilon h_M}(\epsilon h_M -z)^2 e^{-2z\sqrt{k_m^2+k_n^2}}\ dz \right. \nonumber
\\
&\hspace{8cm} \left.+2(k_m^2+k_n^2) \int_0^{\epsilon h_M}(\epsilon h_M -z)^2 z^2 e^{-2z\sqrt{k_m^2+k_n^2}}\ dz \right] \label{Eq}
\end{align}
Now we bound the two $z$ integrals appearing in the last expression.  One can show by basic calculus that for any $z\ge0$,
\begin{equation}\label{max}
e^{-2z\sqrt{k_m^2+k_n^2}}\le 1 \ \ \ \text{and} \ \ z^2e^{-2z\sqrt{k_m^2+k_n^2}}\le\frac{4}{e^2(k_m^2+k_n^2)}.
\end{equation}
Hence,
\[\int_0^{\epsilon h_M}(\epsilon h_M -z)^2 e^{-2z\sqrt{k_m^2+k_n^2}}\ dz\le\int_0^{\epsilon h_M}(\epsilon h_M -z)^2\ dz=\frac{\epsilon^3 h_M^3}{3}\]
and
\[\int_0^{\epsilon h_M}(\epsilon h_M -z)^2 z^2 e^{-2z\sqrt{k_m^2+k_n^2}}\ dz\le\frac{4}{e^2(k_m^2+k_n^2)}\int_0^{\epsilon h_M}(\epsilon h_M -z)^2\ dz=\frac{4\epsilon^3h_M^3}{3e^2(k_m^2+k_n^2)}.\]
Substituting these bounds in Eq (\ref{Eq}), we obtain
\begin{align}
\int_0^{\epsilon h_M}(\epsilon h_M-z)\left\|\deriv{^2u_1}{z^2}  \right\|_{x,y}^2 dz\le \epsilon^3 h_M^3 |\taub|^2 K_1^2\sum_{(m,n)\neq\vec{0}}(k_m^4+k_n^4+2k_m^2k_n^2) \left|\hat{h}(m,n)\right|^2 \label{Eq2}
\end{align}
where we define $K_1\equiv\sqrt{2\alpha(1+8/e^2) /3 }$.  Now, observe that
\begin{align}
\sum_{(m,n)\neq\vec{0}}(k_m^4+&k_n^4+2k_m^2k_n^2) \left|\hat{h}(m,n)\right|^2 \nonumber
\\
&=\sum_{(m,n)\neq\vec{0}}k_m^4 \left|\hat{h}(m,n)\right|^2 +\sum_{(m,n)\neq\vec{0}}k_n^4\left|\hat{h}(m,n)\right|^2 +2\sum_{(m,n)\neq\vec{0}}k_m^2k_n^2\left|\hat{h}(m,n)\right|^2 \nonumber
\\
&=\sum_{(m,n)\neq\vec{0}}\left|-k_m^2 \hat{h}(m,n)\right|^2 +\sum_{(m,n)\neq\vec{0}}\left|-k_n^2\hat{h}(m,n)\right|^2 +2\sum_{(m,n)\neq\vec{0}}\left|-k_mk_n\hat{h}(m,n)\right|^2 \nonumber
\\
&=\left\|\sum_{(m,n)\neq\vec{0}}-k_m^2 \hat{h}(m,n)e^{i(k_mx+k_ny)}\right\|^2_{x,y} +\left\|\sum_{(m,n)\neq\vec{0}}-k_n^2\hat{h}(m,n)e^{i(k_mx+k_ny)}\right\|^2_{x,y} \nonumber
\\
& \ \ \ \ \ \ \ \ \ \ \ \ \ \ \ \ \ \ \ \ \ \ \ \ \ \ +2\left\|\sum_{(m,n)\neq\vec{0}}-k_mk_n\hat{h}(m,n)e^{i(k_mx+k_ny)}\right\|^2_{x,y} \nonumber
\\
&=\left\|\deriv{^2h(x,y)}{x^2}\right\|_{x,y}^2 + \left\|\deriv{^2 h(x,y)}{y^2}\right\|_{x,y}^2 + 2\left\|\deriv{^2 h(x,y)}{x\partial y}\right\|_{x,y}^2 \label{Eq3}
\end{align}
Parseval's Theorem is invoked in the penultimate line to swap each discrete sum for the function norm of a Fourier series. Substituting Eq \ref{Eq3} into Eq \ref{Eq2}, we may rewrite Eq \ref{integ} as
\begin{align*}
R^{(1)}&\le\epsilon^2h_M^2K_1|\taub|\sqrt{\left\|\deriv{^2h(x,y)}{x^2}\right\|_{x,y}^2 + \left\|\deriv{^2 h(x,y)}{y^2}\right\|_{x,y}^2 + 2\left\|\deriv{^2 h(x,y)}{x\partial y}\right\|_{x,y}^2}
\\
&=\epsilon^2h_M^2K_1|\taub| \left\| \nabla \nabla h(x,y)\right\|_{x,y}
\end{align*}
giving us our final result.

\subsection{The error $R^{(2)}$}

To compute the second error term $R^{(2)}$, one must have the full solution for the second-order velocity $\vec{u}_2(x,y,z)$. Let $C_u, \ C_v,$ and $C_w$ represent the non-constant-term Fourier coefficients for $\vec{u}_2(x,y,0)$, obtainable via convolution.  That is,
\begin{align}\label{u2bc}
u_2(x,y,0)=-h(x,y)\left.\deriv{u_1}{z}\right|_{z=0}&=-\vec{\hat{x}}\cdot\tilde{\vec{M}}\cdot\taub+\sum_{(m,n)\neq\vec{0}}C_u(m,n)e^{i(k_mx+k_ny)}
\\
v_2(x,y,0)=-h(x,y)\left.\deriv{v_1}{z}\right|_{z=0}&=-\vec{\hat{y}}\cdot\tilde{\vec{M}}\cdot\taub+\sum_{(m,n)\neq\vec{0}}C_v(m,n)e^{i(k_mx+k_ny)}
\\
w_2(x,y,0)=-h(x,y)\left.\deriv{w_1}{z}\right|_{z=0}&=\sum_{(m,n)\neq\vec{0}}C_w(m,n)e^{i(k_mx+k_ny)}
\end{align}
Now, inserting the boundary conditions into the general solution form, one obtains:
\begin{align}
u_2 &=-\vec{\hat{x}}\cdot\tilde{\vec{M}}\cdot\taub+\sum_{(m,n)\neq\vec{0}}e^{-z\sqrt{k_m^2+k_n^2}}\left(\frac{-k_mk_n C_v-k_m^2C_u}{\sqrt{k_m^2+k_n^2}}z - i mz C_w+C_u\right)e^{i(k_mx+k_ny)}
\\
v_2 &=-\vec{\hat{y}}\cdot\tilde{\vec{M}}\cdot\taub+\sum_{(m,n)\neq\vec{0}}e^{-z\sqrt{k_m^2+k_n^2}}\left(\frac{-k_mk_n C_u-k_n^2C_v}{\sqrt{k_m^2+k_n^2}}z - i nz C_w+C_v\right)e^{i(k_mx+k_ny)}
\\
w_2&=\sum_{(m,n)\neq\vec{0}}e^{-z\sqrt{k_m^2+k_n^2}}\left((\sqrt{k_m^2+k_n^2}C_w - imC_u-inC_v)z+C_w\right)e^{i(k_mx+k_ny)}
\end{align}

The derivation of the bound on $R^{(2)}$ follows a similar sequence as the proof in the last subsection for $R^{(1)}$.  Steps will be abbreviated accordingly.  First, the Cauchy-Schwartz inequality allows us to bound the integral:

\begin{align*}
\left|\int_0^{\epsilon h(x,y)}\deriv{\vec{u}_2}{z}\ dz \right| & \le  \sqrt{\epsilon h_M}\sqrt{\int_0^{\epsilon h_M} \left| \deriv{\vec{u}_2}{z}\right|^2 dz} 
\end{align*}
As in the previous section, the next step is to insert this result into the definition of $R^{(2)}$ and expand the function norm, ultimately giving us
\begin{align}\label{r2integ}
R^{(2)} &\le\sqrt{\epsilon h_M\int_0^{\epsilon h_M}\left\|\deriv{^2 \vec{u}_2}{z}  \right\|_{x,y}^2 dz}. 
\end{align}
Taking the $z$ derivative of  $\vec{u}_2$ above, we find that
\begin{align*}
\left|\deriv{\vec{u}_2}{z}\right|^2=\sum_{j=\{1,2,3\}} \left(\sum_{(m,n)\neq\vec{0}} \right.\vec{c}_{m,n}\cdot\left(\vec{q}_{m,n}\cdot\vec{D}_{(j)}\right.\cdot \vec{p}_{m,n}\ , & \ \vec{q}_{m,n}\cdot\vec{E}_{(j)}\cdot \vec{p}_{m,n}\ , 
\\
& \left. \left. \vec{q}_{m,n}\cdot\vec{F}_{(j)}\cdot \vec{p}_{m,n}     \right) \frac{e^{-z\sqrt{k_m^2+k_n^2}}}{\sqrt{k_m^2+k_n^2}}e^{i(k_m x + k_ny)} \right)^2
\end{align*}
where the vectors $\vec{p}_{m,n}$ and $\vec{q}_{m,n}$ are defined in Eqs \ref{pq}, $\vec{c}_{m,n}=(C_u,C_v,C_w)$, and the matrices $\vec{D}_{(j)}$, $\vec{E}_{(j)}$, and $\vec{F}_{(j)}$ are constant, order one, $4\times5$ matrices for each $j$.

As was done before, we next apply Parseval's Theorem to this result, giving
\begin{align*}
\left\|\deriv{\vec{u}_2}{z}\right\|_{x,y}^2=\sum_{j=\{1,2,3\}} \sum_{(m,n)\neq\vec{0}} \left|\vec{c}_{m,n}\cdot\left(\vec{q}_{m,n}\cdot\vec{D}_{(j)}\cdot \vec{p}_{m,n}\ ,  \ \vec{q}_{m,n}\cdot\vec{E}_{(j)}\cdot \vec{p}_{m,n}\ , \ \vec{q}_{m,n}\cdot\vec{F}_{(j)}\cdot \vec{p}_{m,n}     \right) \frac{e^{-z\sqrt{k_m^2+k_n^2}}}{\sqrt{k_m^2+k_n^2}} \right|^2
\end{align*}
We apply the Cauchy-Schwartz inequality to this result, and simplify/expand algebraically:
\begin{align*}
\left\|\deriv{\vec{u}_2}{z}\right\|_{x,y}^2 &\le \sum_{(m,n)\neq\vec{0}} \beta\ |\vec{c}_{m,n}|^2 |\vec{p}_{m,n}|^2 |\vec{q}_{m,n}|^2 \frac{e^{-2z\sqrt{k_m^2+k_n^2}}}{k_m^2+k_n^2}
\\
&=\sum_{(m,n)\neq\vec{0}} \beta\ |\vec{c}_{m,n}|^2 |\vec{p}_{m,n}|^2 \frac{2k_m^4+2k_n^4+4k_m^2k_n^2}{k_m^2+k_n^2}e^{-2z\sqrt{k_m^2+k_n^2}}
\\
&= \sum_{(m,n)\neq\vec{0}} 2\beta\ |\vec{c}_{m,n}|^2 |\vec{p}_{m,n}|^2 (k_m^2+k_n^2)e^{-2z\sqrt{k_m^2+k_n^2}}
\\
&= \sum_{(m,n)\neq\vec{0}} 2\beta\ |\vec{c}_{m,n}|^2 (1+2z^2(k_m^2+k_n^2)) (k_m^2+k_n^2)e^{-2z\sqrt{k_m^2+k_n^2}}
\end{align*}
where
\[\beta=\sum_{j=\{1,2,3\}} \left|\vec{D}_{(j)}\right|^2+\left|\vec{E}_{(j)}\right|^2+\left|\vec{F}_{(j)}\right|^2.\]

\noindent The integral in Eq \ref{r2integ} can now be bounded:
\begin{align*}
\int_0^{\epsilon h_M} \left\|\deriv{\vec{u}_2}{z}\right\|_{x,y}^2 dz & \le \sum_{(m,n)\neq\vec{0}} 2\beta |\vec{c}_{m,n}|^2  (k_m^2+k_n^2)\left[\int_{0}^{\epsilon h_M} \right. e^{-2z\sqrt{k_m^2+k_n^2}}dz  +   2(k_m^2+k_n^2)\left.\int_0^{\epsilon h_M}z^2e^{-2z\sqrt{k_m^2+k_n^2}}dz\right]
\\
&\le \epsilon h_M K_2^2\sum_{(m,n)\neq\vec{0}}  |\vec{c}_{m,n}|^2  (k_m^2+k_n^2)
\end{align*}
where $K_2\equiv\sqrt{2\beta(1+8/e^2)}$ is obtained by bounding the two integrands per  (\ref{max}).

Substituting this bound into Eq \ref{r2integ}, apply Parseval's Theorem again, keeping in mind the definition of $\vec{c}_{m,n}=(C_u,C_v,C_w)$ from Eqs \ref{u2bc}: 
\begin{align}
R^{(2)} &\le\epsilon h_M K_2 \sqrt{\sum_{(m,n)\neq\vec{0}}  |\vec{c}_{m,n}|^2  (k_m^2+k_n^2)}=\epsilon h_M K_2 \sqrt{\sum_{(m,n)\neq\vec{0}}  |i k_m \vec{c}_{m,n}|^2 + \sum_{(m,n)\neq\vec{0}}|i k_n \vec{c}_{m,n}|^2} \nonumber
\\
&=\epsilon h_M K_2 \sqrt{\left\|\deriv{}{x}\left(-h(x,y)\left.\deriv{\vec{u}_1 }{z}\right|_{z=0}\right)\right\|_{x,y}^2  + \left\|\deriv{}{y}\left(-h(x,y)\left.\deriv{\vec{u}_1 }{z}\right|_{z=0}\right)\right\|_{x,y}^2} \nonumber
\\
&=\epsilon h_M K_2 \left\| \nabla \left(h(x,y) \left.\deriv{\vec{u}_1 }{z}\right|_{z=0}\right) \right\|_{x,y}=\epsilon h_M K_2 \left\|\left(\nabla h(x,y)\right) \left.\deriv{\vec{u}_1 }{z}\right|_{z=0}+h(x,y) \left(\nabla\left.\deriv{\vec{u}_1 }{z}\right|_{z=0}\right)\right\|_{x,y} \nonumber
\\
&\le \epsilon h_M K_2 \left(|\nabla h|_M\ \left\| \left.\deriv{\vec{u}_1 }{z}\right|_{z=0}\right\|_{x,y}+ h_M\ \left\|\nabla \left.\deriv{\vec{u}_1 }{z}\right|_{z=0}\right\|_{x,y}\right) \label{R2boundnew}
\end{align}
Computing $|\partial \vec{u_1}/\partial z|^2$ at $z=0$, the solution takes the form:
\begin{align*}
\left|\left.\deriv{\vec{u}_1}{z}\right|_{z=0}\right|^2=\sum_{j=\{1,2,3\}}\left(\sum_{(m,n)\neq\vec{0}} \taub\cdot\left(\vec{s}_{m,n}\cdot\vec{G}_{(j)}\cdot\vec{t}_{m,n} \ , \  \vec{s}_{m,n}\cdot\vec{H}_{(j)}\cdot\vec{t}_{m,n}\right)\hat{h}(m,n)e^{i(k_mx+k_ny)}\right)^2
\end{align*}
For
\[\vec{s}_{m,n}=\left(1,\frac{k_m}{\sqrt{k_m^2+k_n^2}},\frac{k_n}{\sqrt{k_m^2+k_n^2}}\right) \ \ \ \text{and}  \ \ \ \vec{t}_{m,n}=(k_m,k_n)\]
and $\vec{G}_{(j)}$ and $\vec{H}_{(j)}$ constant, dimensionless, order one, $3\times2$ matrices.  Parseval's Theorem and the Cauchy-Schwartz inequality then give:

\begin{align}
\left\|\left.\deriv{\vec{u}_1}{z}\right|_{z=0}\right\|_{x,y}^2 &\le \sum_{(m,n)\neq\vec{0}}\gamma\ |\taub|^2|\vec{s}_{m,n}|^2|\vec{t}_{m,n}|^2\left|\hat{h}(m,n)\right|^2 \nonumber
\\
&=2\gamma\ |\taub|^2  \sum_{(m,n)\neq\vec{0}} (k_m^2+k_n^2)\left|\hat{h}(m,n)\right|^2 \nonumber
\\
&=2\gamma\ |\taub|^2 \left( \left\|\deriv{h}{x}\right\|_{x,y}^2 +  \left\|\deriv{h}{y}\right\|_{x,y}^2 \right) =2\gamma\ |\taub|^2 \left\| \nabla h\right\|^2_{x,y} \ \label{u1z0norm}
\end{align}
for
\[\gamma=\sum_{j=\{1,2,3\}}|\vec{G}_{(j)}|^2+|\vec{H}_{(j)}|^2.\]
By similar means, we can show
\begin{align}
\left\|\nabla\left.\deriv{\vec{u}_1}{z}\right|_{z=0}\right\|^2 &\le \sum_{(m,n)\neq\vec{0}}\gamma\ (k_m^2+k_n^2)|\taub|^2|\vec{s}_{m,n}|^2|\vec{t}_{m,n}|^2\left|\hat{h}(m,n)\right|^2 \nonumber
\\
&=2\gamma\ |\taub|^2  \sum_{(m,n)\neq\vec{0}} (k_m^4+k_n^4+2k_m^2k_n^2)\left|\hat{h}(m,n)\right|^2 \nonumber
\\
&=2\gamma\ |\taub|^2 \left( \left\|\deriv{^2h}{x^2}\right\|_{x,y}^2 +  \left\|\deriv{^2h}{y^2}\right\|_{x,y}^2 + 2\left\|\deriv{^2h}{x\partial y}\right\|_{x,y}^2 \right)= 2\gamma\ |\taub|^2 \left\|\nabla \nabla h\right\|_{x,y}^2\label{delu1z0norm}
\end{align}
Applying (\ref{u1z0norm}) and (\ref{delu1z0norm}) to (\ref{R2boundnew}), and absorbing $\sqrt{2\gamma}$ into the definition of $K_2$, we obtain the final result:
\[R^{(2)}\le \epsilon h_M K_2 |\taub|\left(|\nabla h|_M  \left\| \nabla h\right\|_{x,y} +h_M\  \left\|\nabla \nabla h\right\|_{x,y}\right)\]

\end{document}